\documentclass[a4paper, 11pt]{article}
\usepackage[normalem]{ulem} 
\usepackage{graphicx}
\usepackage[usenames]{color}
\usepackage[flushleft]{threeparttable}
\DeclareGraphicsExtensions{.pdf,.png,.jpg,.mps,.eps,.ps}
\usepackage{amsmath,amssymb,gensymb}
\usepackage[authoryear, round]{natbib}
\usepackage{color}
\usepackage{xcolor}
\usepackage{lscape}
\usepackage{psfrag}
\usepackage[colorlinks,citecolor=blue,urlcolor=blue,bookmarks=false,hypertexnames=true,pagebackref=true]{hyperref} 
\renewcommand{\arraystretch}{2}

\newcommand\nustar{{\it NuSTAR}}

\newcommand\kms{\ifmmode {\rm~km\ s}^{-1} \else ~km s$^{-1}$\fi}
\newcommand\Hunit{\ifmmode {\rm~km\ s}^{-1}\ {\rm Mpc}^{-1}
	\else ~km s$^{-1}$ Mpc$^{-1}$\fi}
\newcommand\ctssec{\ifmmode {\rm~count\ s}^{-1} \else ~count s$^{-1}$\fi}
\newcommand\ergsec{\ifmmode {\rm~erg\ s}^{-1} \else
	~erg s$^{-1}$\fi}
\newcommand\funit{\ifmmode {\rm~erg\ s}^{-1}\;{\rm cm}^{-2} \else
	~ergs s$^{-1}$ cm$^{-2}$\fi}
\newcommand\phflux{\ifmmode {\rm~photon\ s}^{-1}\;{\rm cm}^{-2}
	\else   ~photon s$^{-1}$ cm$^{-2}$\fi}
\newcommand\efluxA{\ifmmode {\rm~erg\ s}^{-1}\;{\rm cm}^{-2}\;{\rm
		\AA}^{-1} \else ~erg s$^{-1}$ cm$^{-2}$ \AA$^{-1}$\fi}
\newcommand\efluxHz{\ifmmode {\rm~erg\ s}^{-1}\;{\rm cm}^{-2}\;{\rm
		Hz}^{-1} \else ~erg s$^{-1}$ cm$^{-2}$ Hz$^{-1}$\fi}
\newcommand\cc{\ifmmode {\rm~cm}^{-3} \else cm$^{-3}$\fi}
\newcommand\FWHM{\ifmmode {\rm~FWHM} \else ${\rm~FWHM}$\fi}
\newcommand\Msun{\ifmmode M_{\odot} \else $M_{\odot}$\fi}
\newcommand\Lsun{\ifmmode L_{\odot} \else $L_{\odot}$\fi}

\newcommand\hbeta{\ifmmode {\rm H}\beta \else H$\beta$\fi}
\newcommand\Kalpha{\ifmmode {\rm K}\alpha \else K$\alpha$\fi}
\newcommand\nh{\ifmmode N_{\rm H} \else N$_{\rm H}$\fi}

\title{Exploring the spectral characteristics of the periodic burster 4U 1323-62: Type-I X-ray burst and persistent emission}
\author{Mahasweta Bhattacharya$^{1}$, Aditya S. Mondal$^{1}\thanks{E-mail: adityas.mondal@visva-bharati.ac.in}$,  Biplab Raychaudhuri$^{1}$, Gulab C. Dewangan$^{2}$  \\
	{\small
		$^{1}$Department of Physics, Visva-Bharati, Santiniketan, West Bengal, 731235, India} \\
	{\small {$^{2}$}Inter-University Centre for  Astronomy \& Astrophysics (IUCAA), Pune, 411007, India}\\
}	

\date{\today}
\begin{document}
	\maketitle
	\begin{abstract}
		We report on the results obtained by the analysis of persistent and type-I thermonuclear X-ray burst emission observed from the periodic burster 4U 1323-62. These analyses are based on the \textit{NuSTAR} observation performed on 2024 August 7 for a total exposure of around 90 ks. The persistent emission is well described by an absorbed thermal Comptonization model. An absorption edge is also detected at an energy of $\sim$ 7.42 keV, which indicates the presence of absorbing material in the vicinity of this system. Six bursts have been observed during this observation, wherein we find the burst recurrence time to be $4.52\pm0.32$ hr. All the bursts exhibit the characteristics of a sharp rise and exponential decay. We perform the time-resolved spectroscopy of the burst spectra described by a model consisting of thermal emission from the neutron star surface and a varying persistent emission component to study the evolution of burst parameters. The enhancement of the persistent emission during burst exposure is characterized by the scaling parameter $f_{\rm a}$, which reflects the increasing strength of the burst–disc interaction with burst intensity, likely driven by Poynting–Robertson drag. The spectral analysis of bursts estimate the average apparent blackbody emitting radius of the neutron star to lie within 1.5–3.5 km. The ignition depths computed from the burst parameters indicate short Type-I thermonuclear bursts from a mixed hydrogen-helium fuel layer.
		
	\end{abstract}
	
	\section{Introduction}
	In a low-mass X-ray binary (LMXB), the neutron star accretes matter from a low-mass companion ($\leq 1M_{\odot}$) via Roche lobe overflow, forming an accretion disk where viscous dissipation heats the inflowing gas \citep{1987huba.conf...35B}. As material accumulates on the stellar surface, unstable nuclear burning ignites once the accreted mass exceeds a critical limit, producing a thermonuclear runaway \citep{2021ASSL..461..209G}. Type-I thermonuclear bursts occur when the X-ray luminosity of a neutron star rises abruptly and decays over a longer timescale \citep{2017A&A...606A.130I}. The burst power generally increases rapidly from the persistent emission level ($\simeq$10$^{36}$ erg s$^{-1}$) to about 10$^{38}$ erg s$^{-1}$ within a few seconds, and subsequently declines exponentially over a timescale of seconds to minutes—exhibiting the typical fast-rise, exponential-decay (FRED) profile \citep{2017MNRAS.468.4735L}.
	
	The accreted hydrogen- and helium-rich material becomes compressed into a dense, thin layer under the neutron star’s intense gravitational field \citep{BERTULANI201656}. Continued accretion raises the temperature and density with increasing depth, aided by heat conduction from the stellar core. As matter accumulates, the base of this layer eventually reaches an ‘ignition depth’ where the temperature and pressure become sufficient to trigger thermonuclear fusion. The resulting energy generation—primarily from the triple-alpha process and the hot CNO cycle—further increases the local temperature, thereby accelerating the fusion rate \citep{kuulkers2003photospheric}.
	
	During a burst, a huge amount of energy ($\sim$10$^{39}$ erg) is released almost instantaneously, causing the heated photosphere to cool over time and producing the characteristic sharp rise and gradual decay in X-ray flux \citep{WIESCHER200751}. The interval between successive bursts, referred to as the recurrence time, is closely associated with the parameter $\alpha$, defined as the ratio of persistent (accretion) to burst emission flux \citep{1993ApJ...413..324T, wang2024burst}. When the energy released by thermonuclear runaway exceeds the neutron star’s gravitational binding energy, radiation pressure drives the photosphere outward, resulting in photospheric radius expansion (PRE). This occurs when the burst luminosity temporarily reaches or surpasses the Eddington limit, altering the local mass accretion rate. As the radiation pressure declines, the accreting matter experiences Poynting–Robertson drag, modifying both the accretion flow and luminosity \citep{1977cami.coll..121S}. Subsequently, the radiation energy dissipates, the photosphere cools and contracts back to the stellar surface \citep{10.1093/mnras/220.2.339}. In the absence of huge radiation pressure, the gravitational pull restores the accretion rate to its pre-burst steady level \citep{Peng_2025}.
	
	The source 4U 1323-62 was first detected by \textit{Uhuru} and \textit{Ariel V} missions in 1970s which were later reported in 1978 and 1981, respectively (\citealt{forman1978fourth}; \citealt{warwick1981ariel}). The source was classified as a neutron star on the first detection of a burst using \textit{EXOSAT} and one hour intensity dips were reported in 1985 (\citealt{1985SSRv...40..287V}). One X-ray burst was reported during the dip in the 2-10 keV energy regime with peak flux $\sim$ 7 $\times$10$^{-10}$ ergs cm$^{-2}$ s$^{-1}$. The source is also known as XB 1323-619 and it has been reported to show intensity dips and X-ray bursts (\citealt{balucinskachurch1999xraystudydippinglow}; \citealt{barnard2001}; \citealt{boirin2005highly}; \citealt{balucinska2009neutral}; \citealt{Bhulla_2020}). During the \textit{BeppoSAX} observation of the source in 1997, \cite{balucinskachurch1999xraystudydippinglow} reported a total of 10 X-ray bursts and 12 intensity dips. Apart from 4U 1826-24, it is the other source that is designated as a 'clocked' burster as their bursting remains periodic over extended periods of time (\citealt{balucinska2009neutral}). The dipping occurs due to absorption of the emitted X-ray by the orbital obscuration of the X-ray emitting region by an absorbing region which is structured azimuthally around the neutron star (\citealt{1982ApJ...253L..61W}). The dipping periodicity indicates that the source is viewed nearly in edge-on orientation ($i \approx 90^\circ$), hence, the inclination angle (angle between the direction of the line-of-sight and rotation axis of the accretion disk) $i \sim 60^\circ - 80^\circ$ (\citealt{1987A&A...178..137F}). \cite{barnard2001} reported on the 1997 observation of both \textit{BeppoSAX} and \textit{RXTE}. For \textit{RXTE} observation, they reported seven bursts and seven dips that repeated almost periodically. The last \textit{AstroSat} observation of 4U 1323-62, during the period 2016-17, reported six bursts with recurrence time $\sim$9400 s and two dips with time separation 11.928 hr (\citealt{Bhulla_2020}). The irregular intensity dips were previously reported with a recurrence time $\sim$ 2.94 hr from the \textit{INTEGRAL} observation with no variation in the dips within 20-40 keV energy regime (\citealt{balucinskachurch1999xraystudydippinglow}). \cite{boirin2005highly} reported presence of Fe \textsc{xxv} and Fe \textsc{xxvi} 1s-2p absorption lines around $\sim$ 6.9 keV which are similar to the values reported from \textit{XMM-Newton} observation by \cite{Church_2005}.
	
	Non-dip spectra was modelled via two components, one of which was absorbed and the other was unabsorbed, it is known as 'absorbed$+$unabsorbed' approach (\citealt{1986ApJ...308..199P}; \citealt{smale1992ginga}). In the 1985 observation by \citealt{1985SSRv...40..287V}, the blackbody component for the burst spectrum was fitted that showed cooling during the burst decay. It was reported to have an overall flux of 7-8 $\times$ 10$^{-11}$ ergs cm$^{-2}$ s$^{-1}$ in the energy range 2-10 keV with a power law index ranging between $1.1 - 1.8$. \cite{1997ApJ...491..388C} showed that in order to fit the spectral evolution for the emission after dipping, the resultant model requires emission of two components to fit the dip states as well as the spectra formed at the non-dipping states. It is shown previously that the best-fit model, called the 'complex continuum' model, consists of a component to fit the Comptonized emission from the accretion disk corona (extended region) and a blackbody component to fit the emission from the neutron star (point-source) (\citealt{balucinskachurch1999xraystudydippinglow}; \citealt{barnard2001}). \cite{balucinskachurch1999xraystudydippinglow} reported that for the broadband study of the source in the 0.1-150 keV energy regime, the non-bursting non-dipping intervals could be modelled with a power law with a photon index $\Gamma$=$1.48 \pm 0.01$, a cut-off energy $44.1_{-4.4}^{+5.1}$ keV. And, the fit for the blackbody emission provided with the blackbody temperature k$T_{\rm bb}=1.77 \pm 0.25$ keV. The fitting of the persistent emission showed that in the 2-10 keV energy range, the blackbody contribution in the total flux is $\sim$ 15\% whereas during dips, the maximum reduction in intensity is $\sim$ 65\%. \citet{barnard2001} reported the blackbody temperature for the \textit{RXTE} observation as k$T_{\rm bb}=1.79 \pm 0.21$ keV and the cut-off power law index, $\Gamma=1.61 \pm 0.04$. In the \textit{RXTE} observation out of the seven bursts, two bursts were reported to be double burst. Later, \citet{BOIRIN20062759} proposed a photo-ionized absorber model rather than individual Gaussian profile to explain the spectral evolution from persistent to dipping LMXB. Frequency-resolved analyses of the source have previously revealed a low-frequency QPO at $\sim$1 Hz \citep{jonker1998persistent, balman2010, Bhulla_2020}.\\
	
	In the present work, we have carried out a detailed persistent spectral analysis and time-resolved spectral analysis of the short successive type-I X bursts observed during the \textit{NuSTAR} observation. This paper is structured as follows. In section 2, we state the observation details and describe data reduction process. Throughout section 3 we discuss the light curves for the observation for the observed bursts and intensity dips. In section 4 we present the adapted models for the persistent and burst emission to carry out the spectral analysis. Additionally, we discuss the time-resolved analysis of three bursts. Section 5 discusses the observations and inferences drawn from the spectral analysis.\\

	\begin{figure*}
		\centering
		\includegraphics[width=0.5\columnwidth]{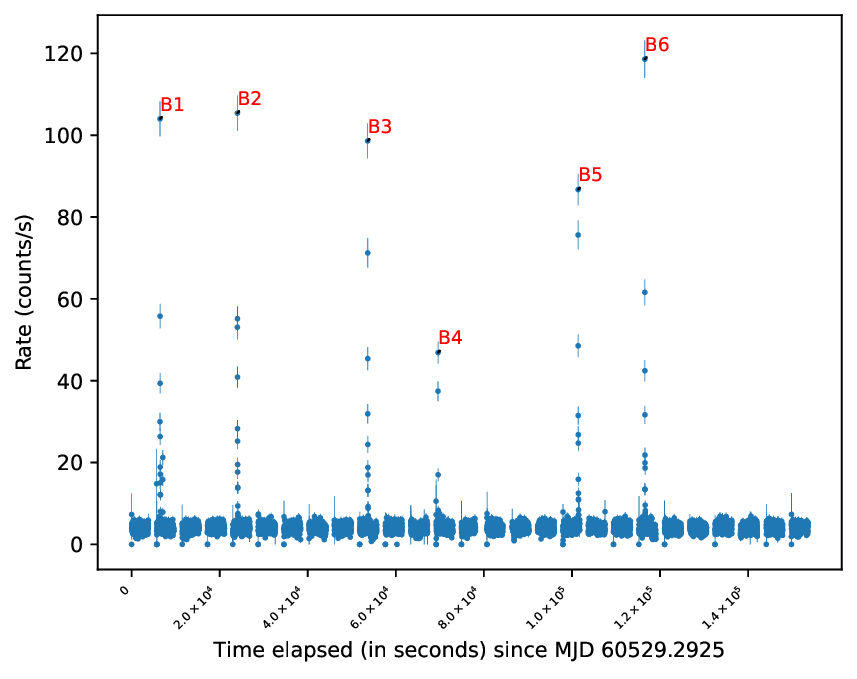}
		\includegraphics[width=0.4\columnwidth]{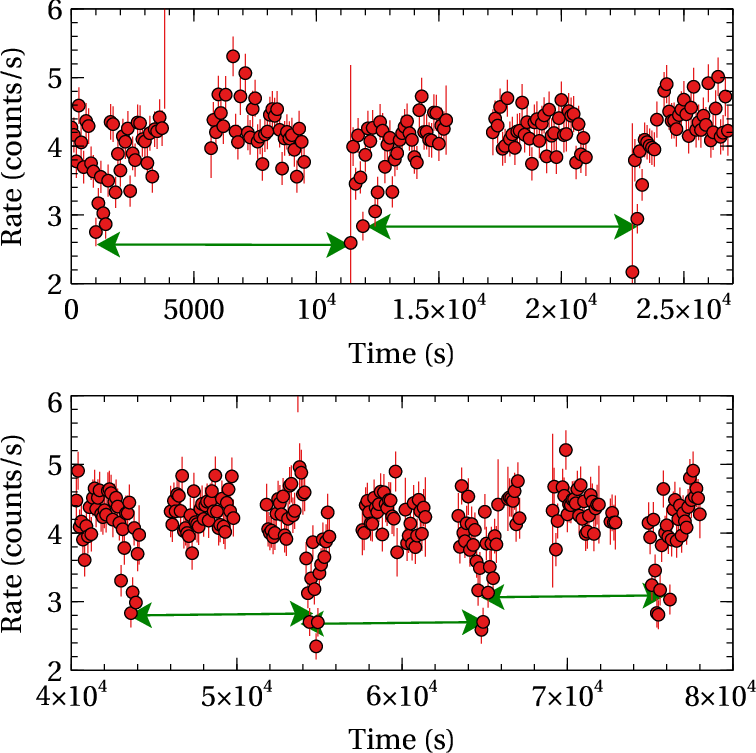}
		\caption{\textit{\textbf{Left panel:}} \nustar{}/FPMA light curve of the source 4U 1323-62 in the energy range 3-79 keV. Entire light curve for the $\sim$90 ks observation with the observed type-I thermonuclear bursts B1, B2, B3, B4, B5 and B6 where the start, end and duration times are noted in Table \ref{tab:Burst_times}; \textit{\textbf{right panel}}: two light curves with 100 s bin size showing regular intensity dips with time interval- 27 ks from the start of observation (\textit{top}), 40 ks within the time segment 40 ks and 80 ks from the beginning of observation (\textit{bottom}), green double headed arrows show the separation between two consecutive intensity dips.}
		\label{fig:entire_lc}
	\end{figure*}
	
	\section{Observation and Data reduction}
	The analysis presented in this work is based on observations of 4U 1323-62 conducted by \textit{NuSTAR} on 2024 August 07 (MJD $60529.2925$) for an effective exposure of $\sim90$ ks under the archival \texttt{Observation ID: 31001021002}. The \textit{Nuclear Spectroscopic Telescopic Array} (\nustar{}), launched by NASA on 2012 June 13, is the first focusing hard X-ray telescope operating in the high energy ($3-79$ keV) range (\citealt{harrison2013nuclear}).\\
	
	The \textit{NuSTAR} data were processed using the data analysis software \texttt{NuSTARDAS v2.1.5}, within \texttt{HEASOFT v6.35.1} employing the latest calibration database (\texttt{CALDB v20250415}). The task \texttt{nupipeline v0.4.12} was used to generate the calibrated and screened event files. A circular extraction region of 120'' radius was used to study the persistent and burst spectrum. From the same chip, the background was selected as a region of radius 120'' far away from the source. HEASOFT provides FTOOLS to deduce the FITS files needed for analysing the observed data. FTOOL \texttt{nuproducts} was used to extract the spectra and light curves from FPMA and FPMB. For time-resolved spectroscopy, good time interval (GTI) files were created separately for the persistent emission and the observed type-I X-ray bursts. During burst analysis, data above 20 keV were excluded due to background dominance.\\

	\section{Light Curve}
	The complete \textit{NuSTAR} FPMA light curve in the 3–79 keV band is shown in Fig. \ref{fig:entire_lc}. The average count rate during the persistent phase was $\sim$ 3-5 counts s$^{-1}$. Overall six type-I thermonuclear bursts were detected with their timing properties listed in Table \ref{tab:Burst_times}. Alongwith the six bursts, seven regular intensity dips have been observed.
	
	We also plotted two light curves with 100 s bin size in the 3–79 keV energy band, using time intervals of the first 27 ks and from 40 ks to 80 ks, which spans a total of 40 ks. Those are presented in the right panel of Fig. \ref{fig:entire_lc}. From the light curves, we found evidence of regular intensity dips with a period of approximately 2.91–2.98 hours (indicated by green double-headed arrows). The source is known for periodic intensity dips, with a period of $\sim$2.94 hours, first discovered through \textit{EXOSAT} observations and later confirmed by other observations (\citealt{1985SSRv...40..287V}; \citealt{boirin2005highly}; \citealt{Bhulla_2020}).
	
\begin{figure*}
		\centering
		\includegraphics[width=0.3\columnwidth]{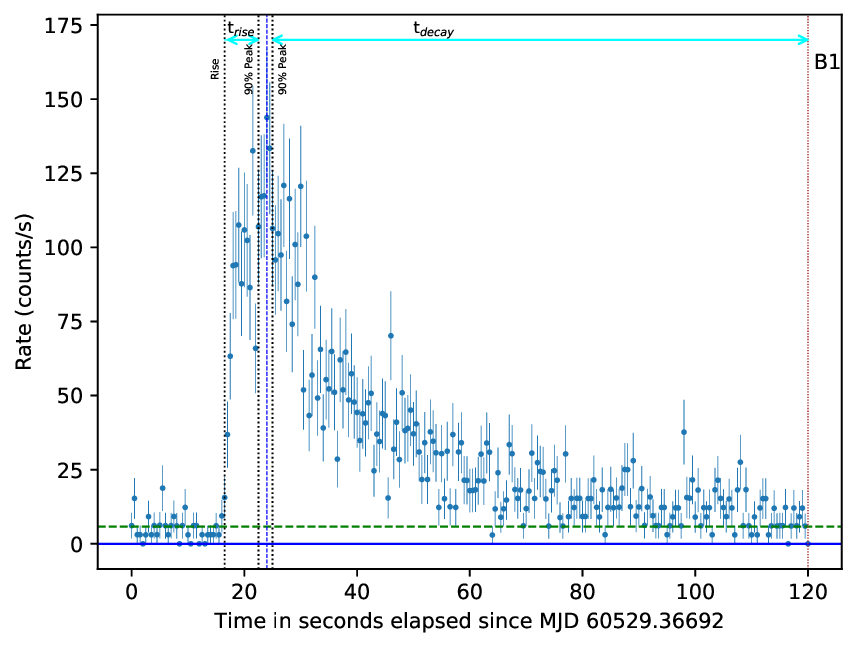}
		\includegraphics[width=0.3\columnwidth]{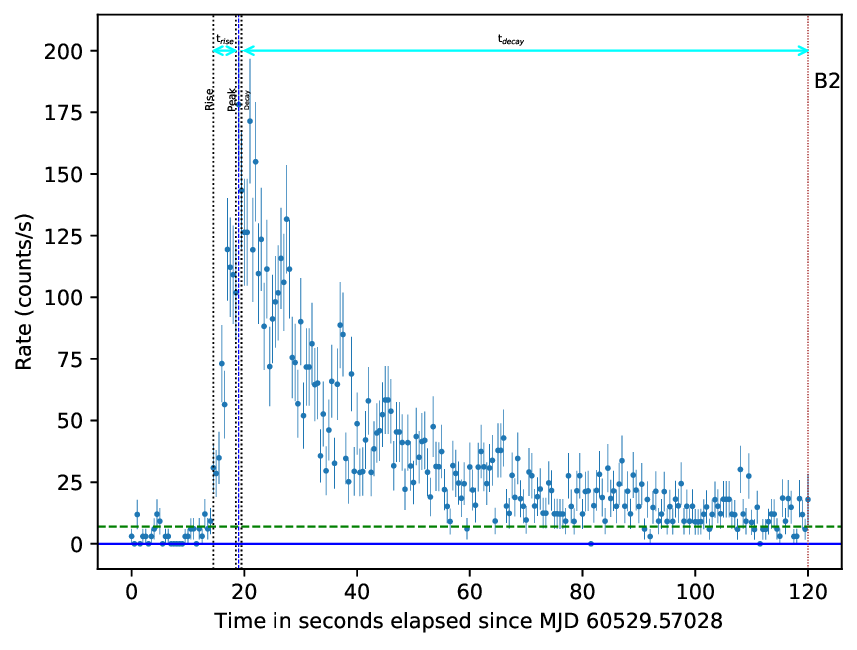}
		\includegraphics[width=0.3\columnwidth]{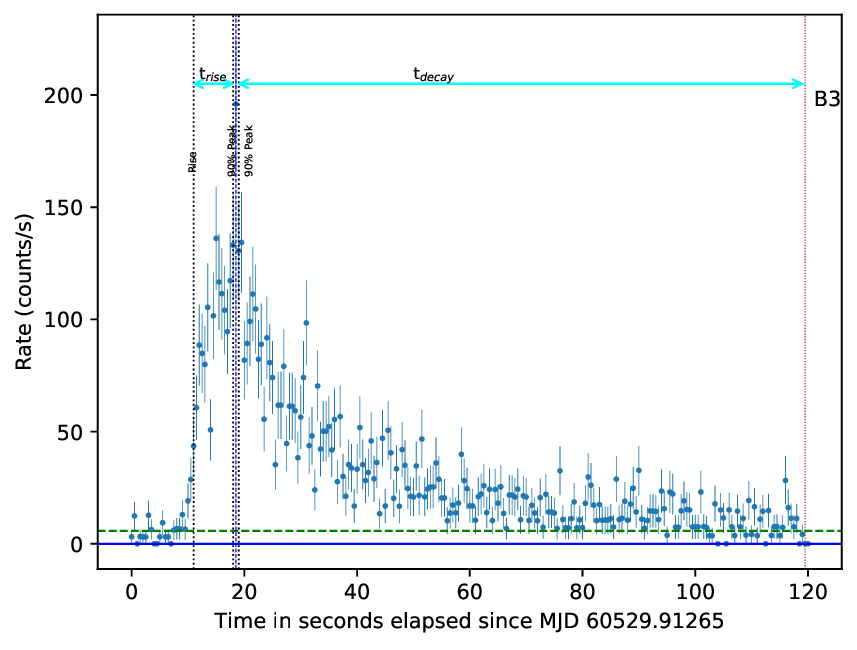}
		\\
		\includegraphics[width=0.3\columnwidth]{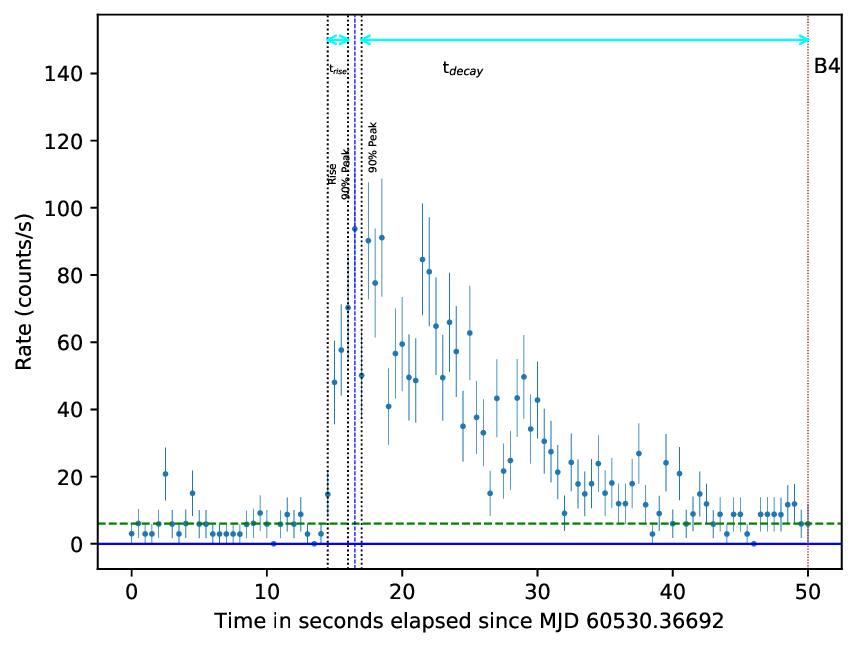}
		\includegraphics[width=0.3\columnwidth]{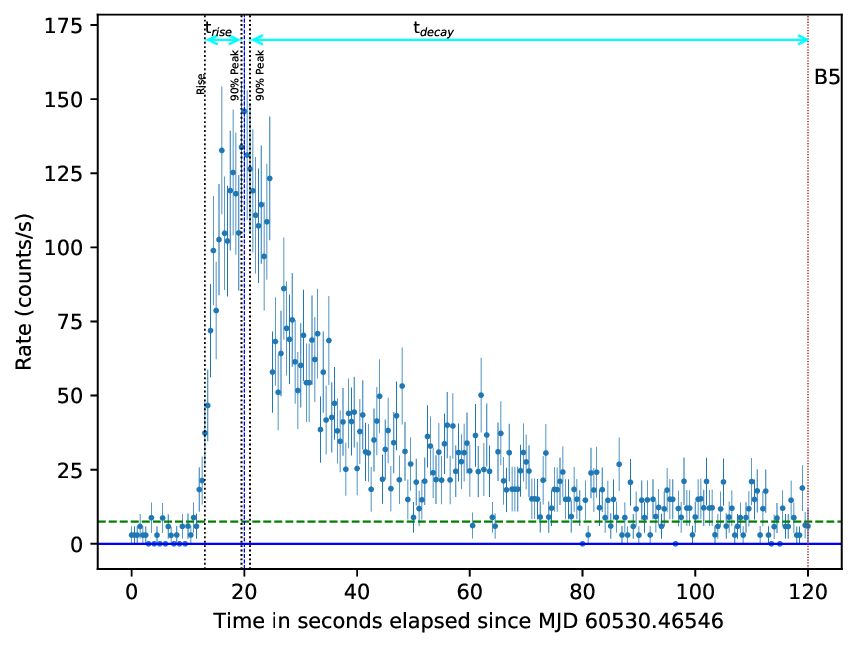}
		\includegraphics[width=0.3\columnwidth]{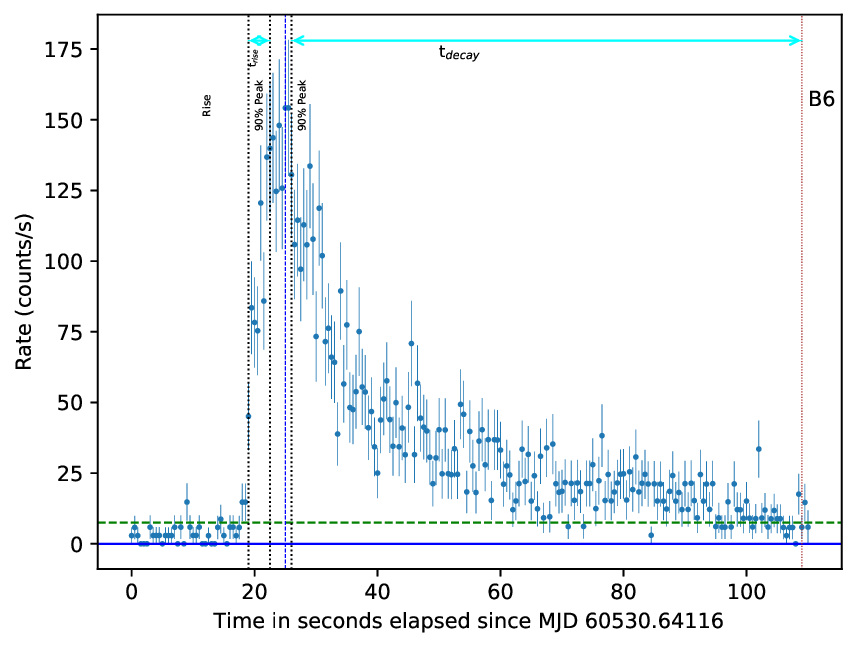}
		\\
		\includegraphics[width=1.0\columnwidth]{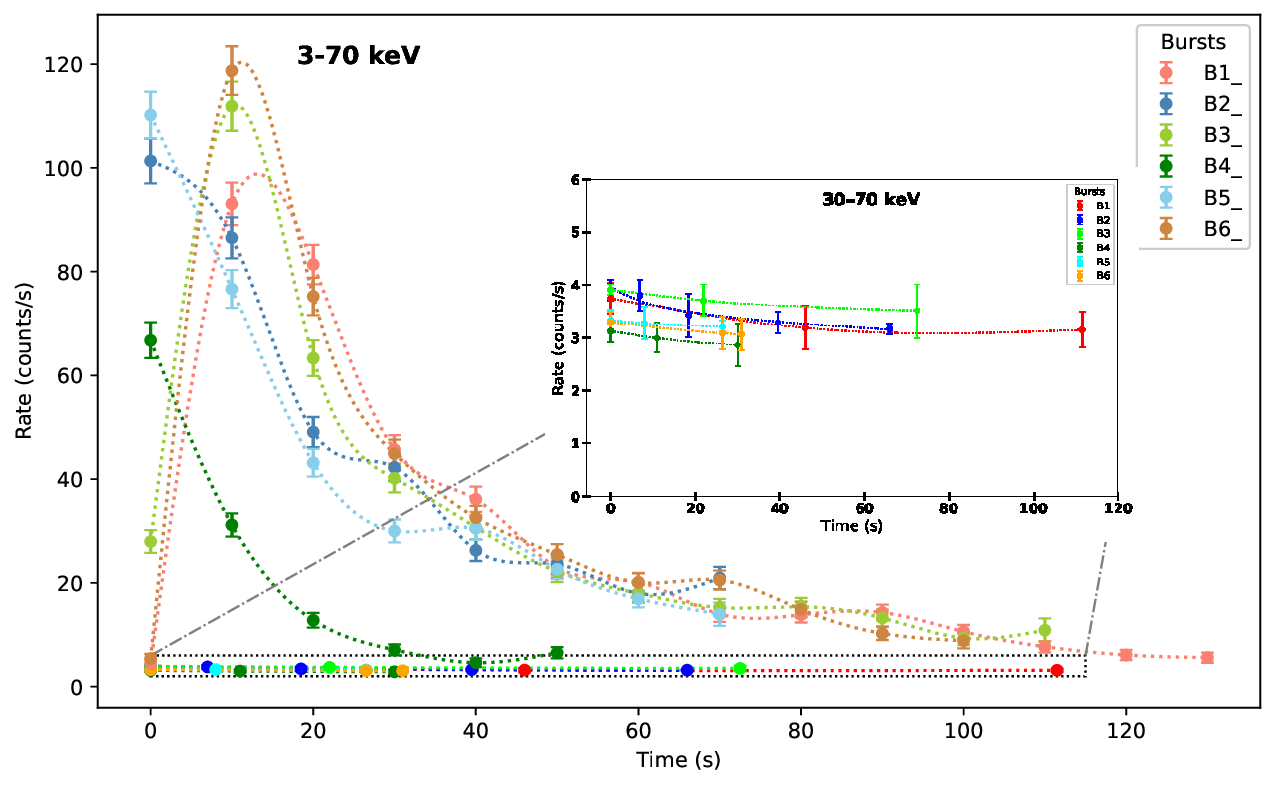}
		
		\caption{\textbf{Upper panel:} 3-79 keV \textit{NuSTAR}/FPMA lightcurve for B1, B2 and B3,respectively. \textbf{Middle panel:}  3-79 keV \textit{NuSTAR}/FPMA lightcurve for B4, B5 and B6, respectively. The horizontal blue line shows the zero count level and the green dotted line for each plot shows the level for 25\% of the peak count rate. Horizontal lines are needed to define $t_{\rm rise}, t_{\rm peak}$ and $t_{\rm decay}$. The vertical blue dash-dotted line corresponds to the peak count rate and brown dashed line marks the end of the exposure interval. \textbf{Lower panel:} \textit{NuSTAR}/FPMA light curve of the six bursts (B1-B6) stacked together aligned at their starting point within 3-70 keV energy regime with a 10 s binsize; \textbf{inset}: 30-70 keV \textit{NuSTAR}/FPMA light curve for B1-B6 with a binsize of 0.5 s, with all bursts aligned at their start.}
		\label{fig:bursts_lc}
	\end{figure*}

	The X-ray bursts as shown in Fig. \ref{fig:bursts_lc} have the characteristic of a fast rise and exponential decay (FRED). The peak count rate during bursts B1, B2, B3, B4, B5 and B6 (in unit of counts s$^{-1}$) are 144, 178, 196, 94, 146 and 154 respectively. The peak count rates are at an average $\sim$25-50 times higher than the persistent level. In Fig. \ref{fig:bursts_lc} for each occasion of burst, we have considered the initiation point of the burst where the count rate is $\sim$25\% of the peak count rate of that particular burst (shown by the green dotted-line). Rise time for the burst is considered as the time taken to increase from the initiation level to the level of 90\% of the peak count rate. The time duration for peak of the burst is considered as the total time for which the count rate remains above 90\% of the peak count rate following \cite{galloway2008thermonuclear}. Decay time for the burst is considered to be the time taken for the count rate to fall down below the initiation level. The total integrated time for each of the bursts is summation of its rise time, peak time and decay time. We have obtained the integrated times for B1, B2, B3, B4, B5 and B6 to be 105 s, 107 s, 109 s, 36 s, 108 s and 91 s, respectively (Table \ref{tab:Burst_times}).\\
	
	The rise times of bursts B1–B6 were measured to be 7 s, 4 s, 7 s, 2 s, 7 s, and 4 s, respectively, while their decay tails were significantly longer—95 s, 102 s, 101 s, 33 s, 99 s, and 83 s. To examine their spectral evolution, we conducted time-resolved spectral analysis for all bursts. Each of the bursts, barring B4 and B6, have exposure above 100 s. For B4 and B6, the burst exposures were shorter (36s and 91s, respectively). We also extracted the light curves for all bursts with a bin size of 0.5 s for the 30-70 keV energy band, starting from the onset of each burst. The light curves in the 3–70 keV (10 s bin size) and 30–70 keV energy ranges are plotted together for comparison and shown in the lower panel of Figure \ref{fig:bursts_lc}. From the inset plot in the lower panel of Fig. \ref{fig:bursts_lc}, the burst count rate in 30-70 keV energy band is approximately 4 counts/s at the peak and decreases to about 3 counts/s during the decay phase of the burst. An exposure of $\sim$ 45 ks, excluding the duration of bursts and dips, was used to extract the persistent spectra.

	\begin{table*}
		\centering
		\caption{Time interval of the bursts with the rise, peak and decay time for the observed bursts B1, B2, B3, B4, B5 and B6 for the \textit{NuSTAR}/FPMA observation of the source 4U 1323-62 in the energy regime 3-79 keV. For each of the bursts, duration of burst exposures are noted with the individual peak counts.}
		\renewcommand{\arraystretch}{0.9}
		\begin{tabular}{ccccccc}
			\hline
			\parbox{0.3cm}{Burst \\ Serial \\ no.} & \parbox{1.5cm}{Start time (s) - End time (s) } & \parbox{0.4cm}{t$_{rise}$ \\ (s)} & \parbox{0.5cm}{t$_{peak}$ \\ (s)} & \parbox{0.6cm}{t$_{decay}$ \\ (s)} & \parbox{1cm}{Burst \\ exposure \\ (s)} & \parbox{1cm}{Peak \\ intensity \\ (counts s$^{-1}$)} \\
			\hline
			B1 & 6430 - 6550 & 7 & 3 & 95 & 105 & 144 \\
			B2 & 24000 - 24120 & 4 & 1 & 102 & 107 & 178 \\
			B3 & 53590 - 53710 & 7 & 1 & 101 & 109 & 196 \\
			B4 & 69570 - 69620 & 2 & 1 & 33 & 36 & 94 \\
			B5 & 101380 - 101500 & 7 & 2 & 99 & 108 & 146 \\
			B6 & 116520 - 116630 & 4 & 4 & 83 & 91 & 154 \\
			\hline
		\end{tabular}
		\label{tab:Burst_times}
	\end{table*}

	\section{Spectral Analysis}
	The spectral analysis for the persistent and burst emission was carried out using \texttt{XSPEC v}12.13.0c (\citealt{1996ASPC..101...17A}). We carried out the time-resolved analysis for three bursts (B1, B2 and B5) having five segments for each in the energy range 3-20 keV. To account for the interstellar absorption along the line of sight, we have used the \texttt{TBabs} model. During the analysis, the neutral hydrogen column depth density is kept constant at $\sim$ $1.2 \times 10^{22}$ $cm^{-2}$ \citep{1985SSRv...40..287V, 1990ARA&A..28..215D}. The source distance is considered to be $\sim$ 10 kpc away \citep{Bhulla_2020, 1995AJ....110.1674C, 1996rftu.proc..141G}.
	
	\subsection{Persistent spectral analysis}
	\label{subsec:pers_time}
	We extracted the persistent spectrum of the source for a $\sim$ 45 ks duration of the current observation using the \textit{NuSTAR} FPMA and FPMB. No X-ray bursts and prominent intensity dips have been observed in this time interval. We fit them simultaneously in the energy band 3 to 60 keV, where the source counts are well above the background counts. We include a constant multiplication factor to account for cross-calibration of the two instruments, FPMA and FPMB.
	
	Initially, we attempted to fit the persistent spectrum with an absorbed cutoff power law model \texttt{(const*Tbabs*cutoffpl)}, which is referred to as Model 1. The Model 1 resulted in $\chi^2 / dof$ = 1222/881 with a photon index of $\Gamma$ = 1.46 $\pm$ 0.01 and the cutoff energy E$_{\rm cut}$ = 72$_{-4}^{+5}$ keV. Previous studies of this source also reported that the energy spectrum is dominated by a power-law component. The measured parameter values, $\Gamma$ and E$_{\rm cut}$, are consistent with values obtained from \textit{BeppoSAX}  and \textit{Suzaku} observations (\citealt{boirin2005highly}; \citealt{balucinska2009neutral}).
	
	To describe the persistent emission more precisely, we employed a thermal Comptonization model, \texttt{nthcomp} (\citealt{1996MNRAS.283..193Z}; \citealt{1999MNRAS.309..561Z}), which involves a seed photon source as a blackbody or a disk-like multi-color blackbody. These seed photons are up-scattered by hot electrons, resulting in the output spectrum. We assumed that the source of the seed photons is the accretion disk ( \texttt{inp$\_$type} = 1). This model \texttt{const*Tbabs*nthcomp} (Model 2) resulted in $\chi^2/dof$ = 1047/880 ($\Delta \chi^{2}$ = -175 for 1 additional parameter, as compared to Model 1) with a photon index of $\Gamma= 1.73\pm 0.01$, the electron temperature kT$_{\rm e}$ = $25_{-2}^{+3}$ keV, and the seed temperature kT$_{\rm bb}$ = 1.33 $\pm$ 0.03 keV. The electron temperature of the corona is comparable to the observed cut-off energy of the spectrum. In addition, we found an absorption feature at $\sim$7-8 keV and tried to fit it with an absorption \texttt{edge} model. The inclusion of the \texttt{edge} model, \texttt{const*Tbabs*edge*nthcomp} (Model 3), improved the fit significantly to  $\chi^2 / dof$ = 928/878 ($\Delta \chi^{2}$ = -119 for 2 additional parameters, compared to Model 2) with edge energy E = 7.42$_{-0.13}^{+0.10}$ keV and optical depth $\tau$ = 0.12 $\pm$ 0.02. The absorption edge detected at 7.42 keV is attributed to K-shell absorption from highly ionized iron in the accretion environment.
	
\begin{figure}
		\centering
		
		\includegraphics[width=0.5\columnwidth,angle=270]{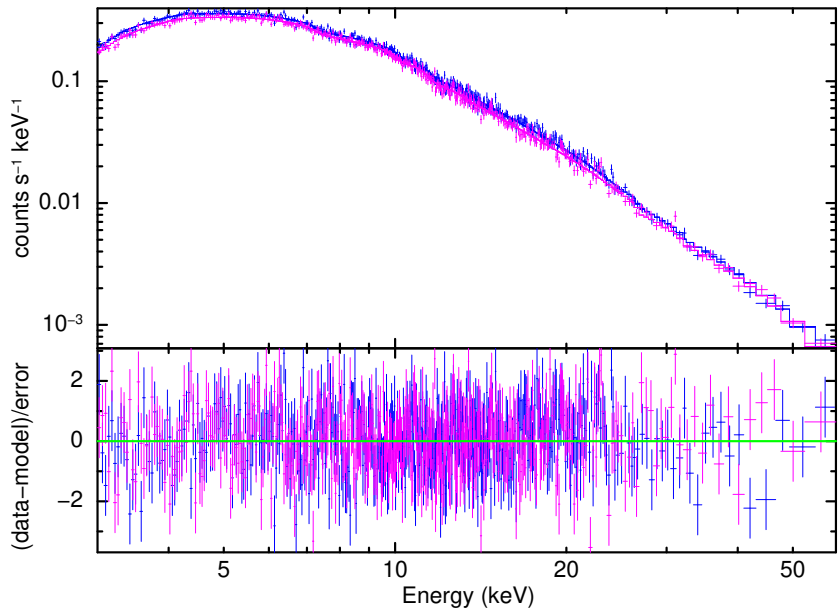}
		\caption{\textbf{Upper Panel:} Unfolded spectral plot for the best-fit data corresponding to the Model 4: \texttt{const*Tbabs*edge*thcomp*diskbb} is shown for the persistent \textit{NuSTAR} spectrum of FPMA (blue) and FPMB (magenta) of the source 4U 1323-62 in the energy range 3-60 keV. \textbf{Lower panel:} Residual plot in 1 $\sigma$ error bar.}
		\label{fig:per_bursts}
	\end{figure}

	To provide a more accurate model description, we further replaced \texttt{nthcomp} with the convolution model \texttt{thcomp} (\citealt{1996MNRAS.283..193Z}; \citealt{2019MNRAS.485.2942N}; \citealt{2020MNRAS.492.5234Z}), an updated version of \texttt{nthcomp} that aligns more closely with the actual Monte Carlo spectra from Comptonization than \texttt{nthcomp} does. The model \texttt{thcomp} involves the upscattering fraction (\textit{f$_{\rm sc}$}) of photons from any seed emission and gives the low-energy cut-off and the high-energy rollover corresponding to the soft tail of the seed photon and electron temperature of the corona, respectively. As we assumed that the accretion disk supplies the seed photons, we applied the model \texttt{const*Tbabs*edge*thcomp*diskbb} (Model 4). This model also satisfactorily described the persistent spectrum, with  $\chi^2/dof$ = 889/877, and provided similar best-fit parameter values, including absorption edge energies compared to Model 3, suggesting that the source geometry is correctly identified.  All the best-fit parameters for the persistent spectral fitting are listed in Table \ref{tab:pers_all}. The persistent spectra for the best fit values corresponding to Model 4 is shown in Fig. \ref{fig:per_bursts}.
	
	We also explored the pre-burst persistent spectra of the source before examining the X-ray burst emissions in detail. For this, we extracted the spectra of the persistent emission from exposures of 600, 600, 1000, 300, 1000, and 1000 s, respectively, just before the onset of bursts B1, B2, B3, B4, B5, and B6, depending on the data availability on the same \textit{NuSTAR} orbit. The 3 – 60 keV energy spectrum of the pre-burst emission was also successfully fitted with an absorbed thermal Comptonization model along with an absorption edge (Model 3). We note that the best-fit parameter values are similar within error when comparing the values obtained from the persistent emission of the $\sim$ 45 ks exposures described earlier. However, parameters such as electron temperature (k$T_{\rm e}$) and edge energy (E$_{\rm edge}$) are not well constrained by the pre-burst persistent spectra due to the limited exposures and low statistics (low count rate) before X-ray bursts. We fixed E$_{\rm edge}$ to $\sim$ 7.42 keV and k$T_{\rm e}$ to $\sim$ 30 keV which are consistent with our earlier findings obtained from the spectral fit of persistent emission using Model 3 (Table \ref{tab:pers_all}). The power-law photon index $\Gamma$, seed photon temperature k$T_{\rm bb}$, absorption optical depth $\tau$, and the corresponding $\chi^2/dof$ values obtained from modeling the pre-burst emissions are reported in Table \ref{tab:pre_burst}.
	
	\begin{table}
		\centering
		\caption{Fit results(Persistent emission): Spectral parameters (best-fit) for the \textit{NuSTAR} observation of the source 4U 1323-62 for the duration of $\sim$ 45 ks, excluding the bursts and dips. Here the model used are given as- Model 1: \texttt{const*Tbabs*cutoffpl} , Model 2: \texttt{const*Tbabs*nthcomp} , Model 3: \texttt{const*Tbabs*edge*nthcomp}  and Model 4: \texttt{const*Tbabs*edge*thcomp*diskbb}}
		\renewcommand{\arraystretch}{0.5}
		
		\begin{tabular}{cccc}
			\hline
			Model no. & Component & Parameter (unit) & Persistent \\
			\hline
			&&&\\
			1 & \textsc{constant} & FPMB (wrt FPMA) & $0.99 \pm 0.01$\\
			& \textsc{Tbabs} & $N_{\rm H} (\times 10^{22})$ $cm^{-2}$ & $1.2$ (f) \\
			& \textsc{cutoffpl} & $\Gamma$ & $1.46 \pm 0.01$ \\
			& & $E_{\rm cut}$ (keV) & $72_{-4}^{+5}$ \\
			& & Norm ($10^{-2}$) & $1.82 \pm 0.03 $ \\
			
			\hline
			
			& $\chi^2$ /$dof$ & & 1222/881 \\
			
			\hline
			&&&\\
			2 & \textsc{nthcomp} & $\Gamma$ & $1.73 \pm 0.01$ \\
			& & $kT_{\rm e}$ (keV) & $25_{-2}^{+3}$ \\
			& & $kT_{\rm bb}$ (keV) & $1.33 \pm 0.03$ \\
			& & Norm ($10^{-3}$) & $9.48 \pm 0.15 $ \\
			
			\hline
			
			& $\chi^2$ /$dof$ & & 1047/880 \\
			\hline
			&&&\\
			3 & \textsc{edge} & E & $7.42_{-0.13}^{+0.10}$ \\
			& & Max $\tau$ & $0.12 \pm 0.02$ \\
			& \textsc{nthcomp} & $\Gamma$ & $1.75 \pm 0.01$ \\
			& & $kT_{\rm e}$ (keV) & $28_{-4}^{+6}$ \\
			& & $kT_{\rm bb}$ (keV) & $1.55 \pm 0.09$ \\
			& & Norm $(10^{-3})$ & $8.79 \pm 0.26$	\\
			\hline
			& $\chi^2$ /$dof$ & & 928/878  \\
			\hline
			&&&\\
			4 & \textsc{edge} & E (keV) & $7.38_{-0.13}^{+0.10}$ \\
			& & Max $\tau$ & $0.10 \pm 0.01$ \\
			& \textsc{thcomp} & $\Gamma_\tau$ & $1.72 \pm 0.01$ \\
			& & $kT_{\rm e}$ (keV) & $18_{-1}^{+2}$ \\
			& & \parbox{2 cm}{Covering \\ fraction (\textit{f$_{\rm sc}$})} & $\ge 0.82$ \\
			& \textsc{diskbb} & $kT_{\rm in}$ (keV) & $1.58 \pm 0.10 $ \\
			& & Norm & $1.12 \pm 0.02$ \\
			\hline
			& $\chi^2$ /$dof$ & & 889/877 \\	
			\hline
			\textit{Note:} & \multicolumn{3}{l}{$dof$ refers to the degrees of freedom ascertained by} \\
			\multicolumn{4}{l}{the components for the particular best-fit model.} \\
		\end{tabular}
		\label{tab:pers_all}
	\end{table}
	
	\begin{table}
		\centering
		\caption{Best-fitted parameters of the pre-burst persistent spectrum for the \textit{NuSTAR} observation of the source 4U 1323-62 in the energy range 3-60 keV from exposures of 600, 600, 1000, 300, 1000, and 1000 s, respectively, just before the onset of bursts B1, B2, B3, B4, B5, and B6. Here the model used is - Model 3: \texttt{const*Tbabs*edge*nthcomp} where the frozen parameters are: for \texttt{edge} component - edge energy, E$_{\rm edge}$= 7.42 keV and electron temperature, k$T_{\rm e}$ = 30 keV.}
		\renewcommand{\arraystretch}{0.5}
		\begin{tabular}{ccccc}
			\hline
			\parbox{1cm}{Before \\ Burst}	& $\Gamma$ & k$T_{bb}$ & $\tau$ & $\chi^2 / dof$ \\
			\hline
			&&&&\\
			1 & 1.74 $\pm$ 0.05 & 1.50 $\pm$ 0.31 & 0.15 $\pm$ 0.09 & 46 / 57 \\
			& & & & \\
			2 & $1.78_{-0.09}^{+0.07}$ & 1.86$\pm$0.5 & $\le 0.19$ & 43 / 49 \\
			& & & & \\
			3 & 1.74 $\pm$ 0.04 & 1.62 $\pm$ 0.21 & 0.19 $\pm$ 0.07 & 103 / 90 \\
			& & & & \\
			4 & 1.77 $\pm$ 0.07 & $1.27_{-0.10}^{+0.45}$ & $\le0.18$ & 31 / 25 \\
			& & & & \\
			5 & 1.72 $\pm$ 0.03 & 1.35 $\pm$ 0.25 & 0.12 $\pm$ 0.05 & 87/ 98 \\
			& & & & \\
			6 & 1.76 $\pm$ 0.05 & 1.79 $\pm$ 0.17 & 0.16 $\pm$ 0.07 & 92 / 90 \\
			\hline
		\end{tabular}
		\label{tab:pre_burst}
	\end{table}

	\subsection{Burst spectral analysis}
	Initially, we extracted the spectra for all the individual X-ray bursts (Table \ref{tab:Burst_times}) and performed simultaneous spectral fitting in the energy range of 3–20 keV. We used the absorbed blackbody model, \texttt{Tbabs*bbodyrad} (Model 5), to fit the burst spectra. The \texttt{bbodyrad} model has two parameters, the blackbody temperature, kT$_{\rm bb}$, and the normalization. We found that the blackbody Model 5 yielded a poor fit of $\chi$$^{2}/dof$ = 646/458. For B1, Model 5 failed to reproduce the burst spectra at both low as well as high energy ranges and for the other bursts, discrepancies appeared mainly at higher energies. To address the observed excesses, we incorporated the thermal Comptonization component \texttt{nthcomp} along with the \texttt{edge} component resulting in composite model \texttt{Tbabs*(edge*nthcomp+bbodyrad)} (Model 6). All the parameters of \texttt{edge} and \texttt{nthcomp} were fixed to their best-fit values corresponding to persistent emission (Table \ref{tab:pers_all}), excluding the parameter \textit{norm} of \texttt{nthcomp}, which was allowed to vary during spectral fit. Model 6 significantly improved the fit, resulting in a $\chi^{2}/dof$ of 530/457 ($\Delta\chi^2 = -136$ for 1 additional parameter, compared to Model 5).\\
	
\begin{figure}
		\centering
		\includegraphics[width=0.5\columnwidth, angle=270]{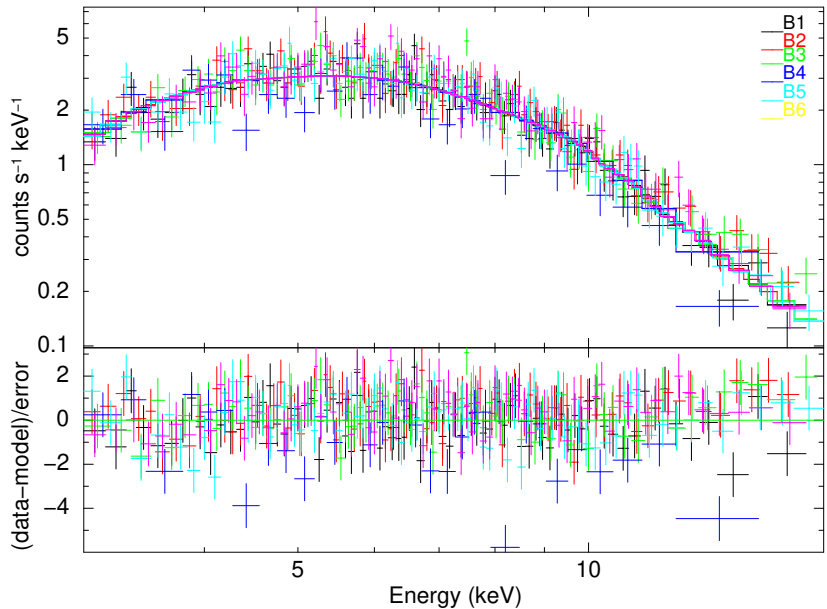}
		\caption{\textbf{Upper panel:} Unfolded spectra of all the bursts are shown for the best-fit data obtained for the model \texttt{Tbabs*(edge*nthcomp+bbodyrad)} as indicated by the legend for colour at the top right corner of the plot for \textit{NuSTAR} telescopic observation of FPMA in the energy range 3-20 keV for the source 4U 1323-62. \textbf{Lower panel:} Residual plot within 1 $\sigma$ error bar for the difference of data vs. model.}
		\label{fig:all_bursts}
	\end{figure}

	The burst spectra are shown in Fig. \ref{fig:all_bursts}. As a best-fit value for all the bursts from Model 6, we obtained the blackbody temperature as k$T_{\rm bb}$ = 1.802 $\pm$ 0.027 keV and radius for the blackbody emitting region $\sim$ $2.38_{-0.06}^{+0.07}$ km. The \texttt{nthcomp} norm is obtained as $0.025 \pm 0.002$. This methodology provides us a suitable overall model for the burst spectral analysis, but there may be a significant change of the physical parameters throughout the time-integrated burst spectra. Each burst spectrum comprises of distinct rise, peak and decay phases. There is a rapid change of the physical parameters during each of the phases, particularly during the burst peak. For instance, the temperature changes significantly from the pre-burst to the burst phase, but such rapid variations are averaged out in the time-integrated spectrum. This motivates us to perform the time-resolved analysis for the bursts in Section \ref{sec:time-resolved} to capture the evolution of the physical parameters. The time-resolved spectral analysis was performed following a similar approach of $f_{\rm a}$ model adopted by \cite{mondal2025nustar} for the “clocked burster” GS 1826–24.\\
	
	We tested other scenarios to fit the burst spectra apart from the $f_{\rm a}$ model. First, we attempted to fit the burst spectra by the model {\tt TBabs*thcomp*bbodyrad}, considering that the burst photons could be affected by the corona/boundary layer \citep{2022ApJ...936...46C}. Here, {\tt thcomp} \citep{2020MNRAS.492.5234Z} is a more accurate version of {\tt nthcomp}, and the parameter values of this model are extracted from the persistent fit. However, this model failed to fit the residuals observed at low and high energies and provided a fit to the burst spectra that was significantly worse compared to the previous $f_{\rm a}$ model. Second, we tried the model {\tt TBabs*(bbodyrad+powerlaw+nthcomp)} to fit the burst spectrum, in which the {\tt nthcomp} parameter values were fixed to the persistent emission \citep{2025ApJ...986...16J}. However, the current data failed to constrain all the free parameters simultaneously and could not produce conclusive results. Third, to fit the burst spectra, another combination of models like {\tt TBabs*(bbodyrad+\\relxillNS+nthcomp)} can be tested in which {\tt relxillNS} accounts for the reflection of the burst photons from the accretion disk \citep{2023A&A...670A..87L}. However, we could not implement this model because of the short exposure time of each segment of the burst spectrum and the lack of a priori knowledge of some important disk-related parameters. The above-mentioned facts restricted us from discussing these models. \\

	\subsection{Time-resolved analysis of the burst spectra}
	\label{sec:time-resolved}
	
	The burst exposures for B1–B6 were divided into shorter time segments to track the evolution of spectral parameters. Except for B4 and B6, each burst lasted over 100 s and was divided into five 20 s segments, with any remaining time beyond 100 s included in the final segment (S5). The segments were labeled sequentially as S1–S5. Due to limited data points and poor spectral fits, B4 and B6 were excluded from the time-resolved analysis. Additionally, B3 was omitted because its later segments (S3–S6) had significantly fewer counts compared to the corresponding intervals in B1, B2, and B5.
	
	We did time-resolved spectroscopy for three of the observed six bursts, namely B1, B2 and B5. Initially, we used the model \texttt{Tbabs*bbodyrad} (Model 5) to fit the burst spectra. The corresponding best-fit spectra for B1, B2 and B5 are presented in the upper panels of Fig. \ref{fig:B1_b,f}, \ref{fig:B2_b,f} and \ref{fig:B5_b,f}, respectively. It is evident that the single-component \texttt{bbodyrad} model does not adequately reproduce the observed spectra, as significant residuals are present in the higher energies. Following the approach of \cite{worpel2013evidence}, the burst spectra were subsequently fitted with a two component model comprising a blackbody component \texttt{bbodyrad} to describe the burst emission and the persistent emission is modeled using \texttt{edge*nthcomp}, scaled by a variable multiplicative factor ($f_{\rm a}$) that represents the enhancement of the persistent emission. The parameters for \texttt{bbodyrad} component and $f_{\rm a}$ were allowed to vary freely. We used the model \texttt{Tbabs*($f_{\rm a}$*edge*nthcomp+bbodyrad)} (Model 7) to analyse the burst emission. A value of $f_{\rm a}=1$ indicates that the persistent emission during the burst is identical to its pre-burst level; while $f_{\rm a}>1$ implies an enhancement of the persistent emission during burst. The \texttt{nthcomp} parameters \texttt{redshift} and \texttt{inp\_type} were set to 0 and 1, respectively, while all other parameters of the \texttt{edge} and \texttt{nthcomp} components were fixed to their best-fit values obtained from the persistent emission fit (see Table \ref{tab:pre_burst}). The unabsorbed fluxes for the total, Comptonized, and blackbody components were derived using the \texttt{XSPEC} convolution model \texttt{cflux} applied to the relevant spectral components.\\
	
	\begin{table*}
		\centering
		\caption{Best-fit results for burst emission B1 (elapsed time stretch 6425 s - 6550 s from the beginning of observation) for the \textit{NuSTAR}/FPMA observation of the source 4U 1323-62 in the energy range 3-20 keV. The model for evaluating the burst emission is Model: 7 \texttt{Tbabs*($f_{\rm a}$ *edge*nthcomp+bbodyrad)}.}
		\renewcommand{\arraystretch}{0.5} 
		\begin{tabular}{ccccccc}
			\hline
			Component & \parbox{1cm}{Parameter \\ (unit)} & S1 (25s) & S2 (20s) & S3 (20s) & S4 (20s) & S5 (40s) \\
			\hline
			\textsc{tbabs} & \textit{$N_{\rm H}$$^*$} & 1.2 (f) & 1.2 (f) & 1.2 (f) & 1.2 (f) & 1.2 (f) \\
			\parbox{1cm}{\textsc{mult. \\ factor}} & $f_{\rm a}$ & $1.78_{-1.63}^{+1.63}$ & $2.56_{-1.78}^{+1.75}$ & $2.14_{-0.87}^{+0.83}$ & $2.45_{-0.63}^{+0.59}$ & $1.66_{-0.39}^{+0.37}$ \\
			& & & & & & \\
			\textsc{cflux} & $F_{\rm f_a*nthcomp} ^\dagger$ & $4.89 \pm 0.16$ &  $ 7.44 \pm 1.29$ & $16.91 \pm 1.43$ & $ 6.97 \pm 1.07$ & $4.87 \pm 0.52$ \\
			& & & & & & \\
			\textsc{bbodyrad} & $kT_{\rm bb}$ (keV) & $2.00_{-0.42}^{+0.49}$ & $2.04_{-0.07}^{+0.07}$ & $1.71_{-0.12}^{+0.12}$ & $1.53_{-0.21}^{+0.21}$ & $1.32_{-0.16}^{+0.17}$ \\
			& & & & & & \\
			& Norm & $1.33_{-1.19}^{+1.35}$ & $13.51_{-1.65}^{+1.79}$ & $9.72_{-2.03}^{+2.68}$ & $5.45_{-2.11}^{+3.74}$ & $6.43_{-2.35}^{+3.94}$ \\
			& & & & & & \\
			& $R_{\rm bb}$ (km) & $1.15_{-0.78}^{+0.49} $ & $3.67_{-0.23}^{+0.24}$ & $3.12_{-0.35}^{+0.40}$ & $2.33_{-0.50}^{0.70}$ & $2.54_{-0.52}^{+0.68}$ \\
			& & & & & & \\
			\textsc{cflux} & $F_{\rm b}^\dagger$ & $2.35_{-0.31}^{+0.32}$ & $22.76_{-0.86}^{+0.86}$ & $3.99_{-0.49}^{+0.49}$ & $2.69_{-0.34}^{+0.35}$ & $1.59_{-0.59}^{+0.19}$ \\
			& & & & & & \\
			\textsc{cflux} & $F_{\rm tot}^\dagger$ & $7.24_{-0.48}^{+0.48}$ & $30.20_{-0.40}^{+0.70}$ & $20.90_{-1.40}^{+0.98}$ & $9.66_{-0.75}^{+0.57}$ & $6.46_{-0.43}^{+0.30}$ \\
			& & & & & & \\
			\hline 
			& $\chi ^2/dof$ & 4.30 / 8 & 46.23 / 41 & 18.15 / 18 & 4.76 / 9 & 14.48 / 12 \\
			
			\hline
			
			\textit{Note:} & \multicolumn{6}{l}{$^{\dagger}$ denotes the  unit of flux in $10^{-10}$ erg cm $^{-2}$ s $^{-1}$.} \\
		\end{tabular}
		\label{tab:fit_B1}
	\end{table*}
	
	\begin{figure*}
		\centering
		
		\includegraphics[width=0.12\columnwidth,angle=270]{fig11_revised.eps}
		\includegraphics[width=0.12\columnwidth,angle=270]{fig12.eps}
		\includegraphics[width=0.12\columnwidth,angle=270]{fig13.eps}
		\includegraphics[width=0.12\columnwidth,angle=270]{fig14.eps}
		\includegraphics[width=0.12\columnwidth,angle=270]{fig15.eps}
		\\
		\includegraphics[width=0.12\columnwidth,angle=270]{fig16_revised.eps}
		\includegraphics[width=0.12\columnwidth,angle=270]{fig17.eps}
		\includegraphics[width=0.12\columnwidth,angle=270]{fig18.eps}
		\includegraphics[width=0.12\columnwidth,angle=270]{fig19.eps}
		\includegraphics[width=0.12\columnwidth,angle=270]{fig20.eps}
		\caption{Unfolded spectra (burst emission) of B1 using \textbf{Upper panel:} Best fit of Model 5: \texttt{Tbabs*bbodyrad}; \textbf{Lower panel:} Best fit of Model 7: \texttt{Tbabs*($f_{\rm a}$*edge*nthcomp+bbodyrad)}, respectively for the \textit{NuSTAR}/FPMA observation in the energy range 3-20 keV of the source 4U 1323-62. The lower plots within each plot shows the residual plot for the particular data and model within 1 $\sigma$ error bar. The time interval corresponding to Burst 1 segments S1-S5 are 25 s, 20 s, 20 s, 20 s and 40 s, respectively.}
		\label{fig:B1_b,f}
	\end{figure*}

	\begin{table*}
		\centering
		\caption{Best-fit results for burst emission B2 (elapsed time stretch 24000 s - 24120 s from the beginning of observation) for the \textit{NuSTAR}/FPMA observation of the source 4U 1323-62 in the energy range 3-20 keV. The model for evaluating the burst emission is Model: 7 \texttt{Tbabs*($f_{\rm a}$*edge*nthcomp+bbodyrad)}.}
		\renewcommand{\arraystretch}{0.5}
		\begin{tabular}{ccccccc}
			\hline
			Component & \parbox{1cm}{Parameter \\ (unit)} & S1 (20s) & S2 (20s) & S3 (20s) & S4 (20s) & S5 (40s) \\
			\hline
			\textsc{tbabs} & \textit{$N_{\rm H}$$^*$} & 1.2 (f) & 1.2 (f) & 1.2 (f) & 1.2 (f) & 1.2 (f) \\
			& & & & & & \\
			\parbox{1cm}{\textsc{mult. \\ factor}} & $f_{\rm a}$ & $2.73_{-1.69}^{+2.03}$ & $6.00_{-2.30}^{+2.28}$ & $3.20_{-1.42}^{+1.32}$ & $1.59_{-0.58}^{+0.65}$ & $2.59_{-0.51}^{+0.43}$ \\
			& & & & & & \\
			\textsc{cflux} & $F_{\rm f_a*nthcomp} ^\dagger$ & $7.82 \pm 1.02$ & $17.70 \pm 2.07$ & $9.27 \pm 0.99$ & $1.13 \pm 1.19$ & $7.54 \pm 0.55$ \\
			& & & & & & \\
			\textsc{bbodyrad} & $kT_{\rm bb}$ (keV) & $2.09_{-0.21}^{+0.22}$ & $2.13_{-0.09}^{+0.09}$ & $1.63_{-0.13}^{+0.12}$ & $1.91_{-0.16}^{+0.15}$ & $1.09_{-0.20}^{+0.20}$ \\
			& & & & & & \\
			& Norm & $2.68_{-1.44}^{+1.57}$ & $9.55_{-1.70}^{+1.82}$ & $10.83_{-2.40}^{+2.97}$ & $4.17_{-1.02}^{+1.27}$ & $13.74_{-6.40}^{+17.71}$ \\
			& & & & & & \\
			& $R_{\rm bb}$ (km) & $1.64_{-0.53}^{+0.42}$ & $3.09_{-0.29}^{+0.28}$ & $3.29_{-0.39}^{+0.42}$ & $2.04_{-0.27}^{-0.29}$ & $3.71_{-1.00}^{+1.34}$ \\
			& & & & & & \\
			\textsc{cflux} & $F_{\rm b}^\dagger$ & $5.06_{-0.48}^{+0.49}$ & $19.45_{-0.91}^{+0.88}$ & $6.95_{-0.49}^{+0.51}$ & $6.28_{-0.43}^{+0.43}$ & $1.37_{-0.20}^{+0.22}$ \\
			& & & & & & \\
			\textsc{cflux} & $F_{\rm tot}^\dagger$ & $12.88_{-0.86}^{+0.70}$ & $37.15_{-1.67}^{+0.87}$ & $16.22_{-0.73}^{+0.76}$ & $7.41_{-0.49}^{+0.53}$ & $8.91_{-0.42}^{+0.59}$ \\
			& & & & & & \\
			\hline
			& $\chi ^2/dof$ & 13.21 / 12 & 33.44 /41 & 15.43 / 18 & 10.95 / 11 & 11.83 / 17 \\
			
			\hline
			
			\textit{Note:} & \multicolumn{6}{l}{$^{\dagger}$ denotes the  unit of flux in $10^{-10}$ erg cm $^{-2}$ s $^{-1}$.} \\
		\end{tabular}
		
		\label{tab:fit_B2}
	\end{table*}

	\begin{figure*}
		\centering
		\includegraphics[width=0.12\columnwidth,angle=270]{fig21.eps}
		\includegraphics[width=0.12\columnwidth,angle=270]{fig22.eps}
		\includegraphics[width=0.12\columnwidth,angle=270]{fig23.eps}
		\includegraphics[width=0.12\columnwidth,angle=270]{fig24.eps}
		\includegraphics[width=0.12\columnwidth,angle=270]{fig25.eps}
		\\
		\includegraphics[width=0.12\columnwidth,angle=270]{fig26.eps}
		\includegraphics[width=0.12\columnwidth,angle=270]{fig27.eps}
		\includegraphics[width=0.12\columnwidth,angle=270]{fig28.eps}
		\includegraphics[width=0.12\columnwidth,angle=270]{fig29.eps}
		\includegraphics[width=0.12\columnwidth,angle=270]{fig30.eps}
		\caption{Unfolded spectra (burst emission) of B2 using \textbf{Upper panel:} Best fit of Model 5: \texttt{Tbabs*bbodyrad}; \textbf{Lower panel:} Best fit of Model 7: \texttt{Tbabs*($f_{\rm a}$*edge*nthcomp+bbodyrad)}, respectively for the \textit{NuSTAR}/FPMA observation in the energy range 3-20 keV of the source 4U 1323-62. The lower plots within each plot shows the residual plot for the particular data and model within 1 $\sigma$ error bar.}
		\label{fig:B2_b,f}
	\end{figure*}
	
	The best-fit parameters of the B1 are shown in Table \ref{tab:fit_B1}, and the corresponding spectra are shown in lower panel of Fig. \ref{fig:B1_b,f}. The blackbody emitting radius of the neutron star reaches a maximum 3.67 km during S2, where the \texttt{bbodyrad} flux peaks at $22.76_{-0.86}^{+0.86}$ $\times10^{-10}$ ergs cm $^{-2}$ s $^{-1}$ and then gradually decreases. However, for some segments (S1 and S3) we obtained a low reduced $\chi^2$ value because of insufficient count rates. The blackbody temperature (k$T_{\rm bb}$) for B1 reaches a maximum at S2 $\sim$2.04$\pm0.07$ keV and then decreases monotonically till S5. The best-fit parameters for time-resolved spectroscopy of B2 are listed in Table \ref{tab:fit_B2} and the corresponding spectra are shown in the lower panel of Fig. \ref{fig:B2_b,f}. The \texttt{bbodyrad} flux reaches its peak value during segment S2, after which the \texttt{bbodyrad} flux steadily decreases through S5. The apparent blackbody emitting radius of the neutron star is estimated around $\sim$3.09 km during S2, increasing slightly to 3.29 km in S3. A larger radius of 3.71 km is obtained for S5; however, this value is poorly constrained due to limited statistics ($\sim$0.696), and thus is not considered reliable. The evolution of the blackbody temperature ($kT_{\rm bb}$) shows a maximum of $2.13\pm0.09$ keV in S2, followed by a gradual decrease toward S5. The best-fit spectral parameters for B5 across its five time segments are presented in Table \ref{tab:fit_B5}, with the corresponding spectra and residuals shown in the lower panel of Fig. \ref{fig:B5_b,f}. The blackbody flux reaches its maximum during segment S2 ($16.39_{-0.79}^{+0.79}$$\times10^{-10}$ erg cm $^{-2}$ s $^{-1}$) and subsequently declines in the later segments. The evolution of the blackbody radius closely follows the flux variation, attaining its largest value of approximately 3.5 km during S2, while the blackbody temperature ($kT_{\mathrm{bb}}$) also attains its highest value at this segment. Across all bursts, the persistent emission shows a pronounced enhancement ($f_{\rm a} \ge 1$), ranging within $\sim$1–4 in B1, $\sim$1–8 in B2, and $\sim$1–7 in B3, suggesting a variable mass accretion rate during the burst evolution \citep{worpel2013evidence}.\\
	
	For the segments B1 S1 and B2 S1, the flux due to persistent emission component (F$_{\rm f_a*nthcomp}$) exceeds the blackbody flux (F$_{\rm b}$) whereas the blackbody flux is expected to dominate until the decay phase of the burst. This may occur due to two possible reasons. Firstly, the time duration of the segments B1 S1 and B2 S1 are larger than the typical rise time of the burst, resulting in the inclusion of the persistent emission effect. If we were able to extract the spectrum for S1 for a time bin size of around 10 s or less than this, we could observe a higher blackbody flux than the persistent one. However, we were unable to perform this because of the limited temporal resolution. Due to the poor counting statistics, for S1 (B1), we have to set the time bin size to 25 s (instead of 20s). Secondly, such behaviour can occur during the sudden onset of a burst, when the ignition may momentarily enhance the accretion rate through Poynting–Robertson drag or due to reprocessing of burst photons in the hot corona. At this early stage, the blackbody component has not yet fully developed, while the persistent Comptonized emission remains comparatively high, leading to a transient phase where the persistent emission exceeds the blackbody flux. Similar short intervals where the persistent flux briefly exceeds the blackbody flux have been reported in previous studies \citep{galloway2008thermonuclear, ji2014hard}. The parameter estimates for some other segments, like B1 S5 and B5 S3 yield low reduced $\chi^2$ values. These low values indicate larger uncertainties in the fits, and the corresponding parameters should therefore be interpreted with caution.\\
	
	\begin{table*}
		\centering
		\caption{Best-fit results for the burst emission B5 (elapsed time stretch 101380 s - 101500 s from the beginning of observation) for the \textit{NuSTAR}/FPMA observation of the source 4U 1323-62 in the energy range 3-20 keV. The model for evaluating the burst emission is Model: 7 \texttt{Tbabs*($f_{\rm a}$*edge*nthcomp+bbodyrad)}.}
		\renewcommand{\arraystretch}{0.5}
		\begin{tabular}{ccccccc}
			\hline
			Component & \parbox{1cm}{Parameter \\ (unit)} & S1 (20s) & S2 (20s) & S3 (20s) & S4 (20s) & S5 (40s) \\
			\hline
			\textsc{tbabs} & \textit{$N_{\rm H}$$^*$} & 1.2 (f) & 1.2 (f) & 1.2 (f) & 1.2 (f) & 1.2 (f) \\
			& & & & & & \\
			\parbox{1cm}{\textsc{mult. \\ factor}} & $f_{\rm a}$ & $2.52_{-1.51}^{+1.57}$ & $4.99_{-2.39}^{+2.36}$ & $0.98_{-0.98}^{+1.69}$ & $2.38_{-0.78}^{+0.75}$ & $1.84_{-0.35}^{+0.31}$ \\
			& & & & & & \\
			\textsc{cflux} & $F_{\rm f_a*nthcomp} ^\dagger$ & $6.95 \pm 1.00$ & $14.51 \pm 1.48$ & $2.89 \pm 0.81$ & $6.87 \pm 0.8$ & $4.81 \pm 0.39$ \\
			& & & & & & \\
			\textsc{bbodyrad} & $kT_{\rm bb}$ (keV) & $1.98_{-0.13}^{+0.12}$ & $1.94_{-0.09}^{+0.08}$ & $1.8_{-0.12}^{+1.11}$ & $1.69_{-0.17}^{+0.17}$ & $1.15_{-0.21}^{+1.36}$ \\
			& & & & & & \\
			& Norm & $5.75_{-1.41}^{+1.51}$ & $12.11_{-2.14}^{+2.27}$ & $8.12_{-1.69}^{+1.83}$ & $4.45_{-1.43}^{+1.92}$ & $6.97_{-3.34}^{+7.96}$ \\
			& & & & & & \\
			& $R_{\rm bb}$ (km) & $2.39_{-0.31}^{+0.30}$ & $3.48_{-0.32}^{+0.31}$ & $2.85_{-0.31}^{+0.30}$ & $2.11_{-0.37}^{+0.41}$ & $2.64_{-0.73}^{+1.22}$ \\
			& & & & & & \\
			\textsc{cflux} & $F_{\rm b}^\dagger$ & $8.54_{-0.54}^{+0.53}$ & $16.39_{-0.79}^{+0.79}$ & $8.07_{-0.48}^{+0.48}$ & $3.41_{-0.38}^{+0.38}$ & $0.91_{-0.17}^{+0.17}$ \\
			& & & & & & \\
			\textsc{cflux} & $F_{\rm tot}^\dagger$ & $15.49_{-0.70}^{+0.73}$ & $30.90_{-0.99}^{+0.72}$ & $10.96_{-0.63}^{+0.52}$ & $10.28_{-0.64}^{+0.68}$ & $5.72_{-0.37}^{+0.35}$ \\
			& & & & & & \\
			\hline
			& $\chi ^2/dof$ & 23.10 / 19 & 36.55 / 36 & 7.26 / 15 & 8.39 / 10 & 12.80 / 10 \\
			\hline
			
			\textit{Note:} & \multicolumn{6}{l}{$^{\dagger}$ denotes the  unit of flux in $10^{-10}$ erg cm $^{-2}$ s $^{-1}$.} \\
		\end{tabular}
		
		\label{tab:fit_B5}
	\end{table*}
	
	\begin{figure*}
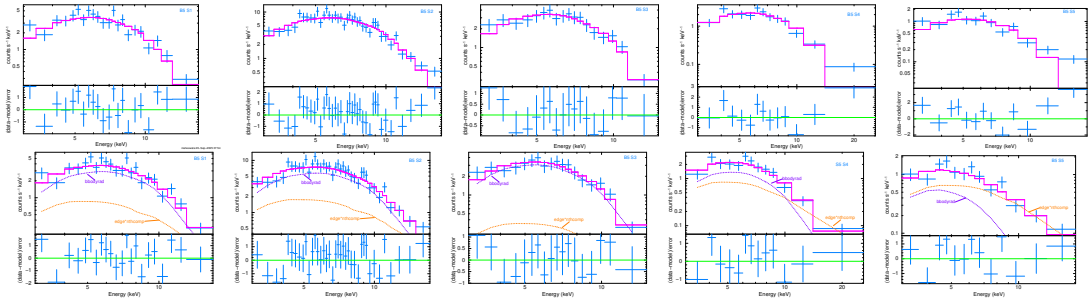

		\centering
		\includegraphics[width=0.12\columnwidth,angle=270]{fig31.eps}
		\includegraphics[width=0.12\columnwidth,angle=270]{fig32.eps}
		\includegraphics[width=0.12\columnwidth,angle=270]{fig33.eps}
		\includegraphics[width=0.12\columnwidth,angle=270]{fig34.eps}
		\includegraphics[width=0.12\columnwidth,angle=270]{fig35.eps}
		\\
		\includegraphics[width=0.12\columnwidth,angle=270]{fig36.eps}
		\includegraphics[width=0.12\columnwidth,angle=270]{fig37.eps}
		\includegraphics[width=0.12\columnwidth,angle=270]{fig38.eps}
		\includegraphics[width=0.12\columnwidth,angle=270]{fig39.eps}
		\includegraphics[width=0.12\columnwidth,angle=270]{fig40.eps}
		\caption{Unfolded spectra (burst emission) of B5 using \textbf{Upper panel:} Best fit data of Model 5: \texttt{Tbabs*bbodyrad}; \textbf{Lower panel:} Best fit of Model 7: \texttt{Tbabs*($f_{\rm a}$*edge*nthcomp+bbodyrad)}, respectively for the \textit{NuSTAR}/FPMA observation in the energy range 3-20 keV of the source 4U 1323-62. The lower plots within each plot shows the residual plot of the particular data and model difference within 1 $\sigma$ error bar.}
		\label{fig:B5_b,f}
	\end{figure*}
	
	\section{Discussion}
	We report on the \textit{NuSTAR} observation of the periodic burster 4U 1323-62, performed on 2024 August 7, for a total exposure of 90 ks. We observed the presence of six thermonuclear X-ray bursts and regular intensity dips in the \textit{NuSTAR} light curve of this source. The observed burst profiles are very similar to those extracted earlier from the \textit{XMM-Newton} EPIC PN and \textit{AstroSat} LAXPC detectors during a persistent interval (\citealt{boirin2005highly}; \citealt{Bhulla_2020}). The observed period ($\sim$ 2.91-2.98 hr) of the intensity dips is also consistent with the other observations (\citealt{1985SSRv...40..287V}; \citealt{boirin2005highly}; \citealt{Bhulla_2020}). During this observation, the persistent emission is well described by the absorbed thermal Comptonization model in the energy band of 3–60 keV. Our persistent spectral analysis suggests that the Comptonization of soft photons originating from the accretion disc is occurring. The power law index $\Gamma$, electron temperature of the corona k$T_{\rm e}$, and the seed photon temperatures kT$_{\rm in}$ obtained from the Comptonization model are 1.72$\pm$ 0.01, 18$_{-1}^{+2}$ keV , and 1.58$\pm$ 0.10 keV, respectively. In addition, we detected an absorption edge at an energy of $\sim$ 7.42 keV, indicating the presence of an absorbing medium in the vicinity of the system. Apart from the dip intervals, the persistent flux remains unchanged on our observed timescale. We measured a 3-60 keV persistent flux of F$_{\rm per} \sim 2.75\times10^{-10}$ erg cm$^{-2}$ s$^{-1}$, which corresponds to a persistent luminosity of L$_{\rm per}$ $\sim$ 3.32$\times$10$^{36}$ erg s$^{-1}$ for an assumed distance of $\sim$ 10 kpc.\\

	\begin{figure*}
		\centering
		\includegraphics[width=0.3\columnwidth]{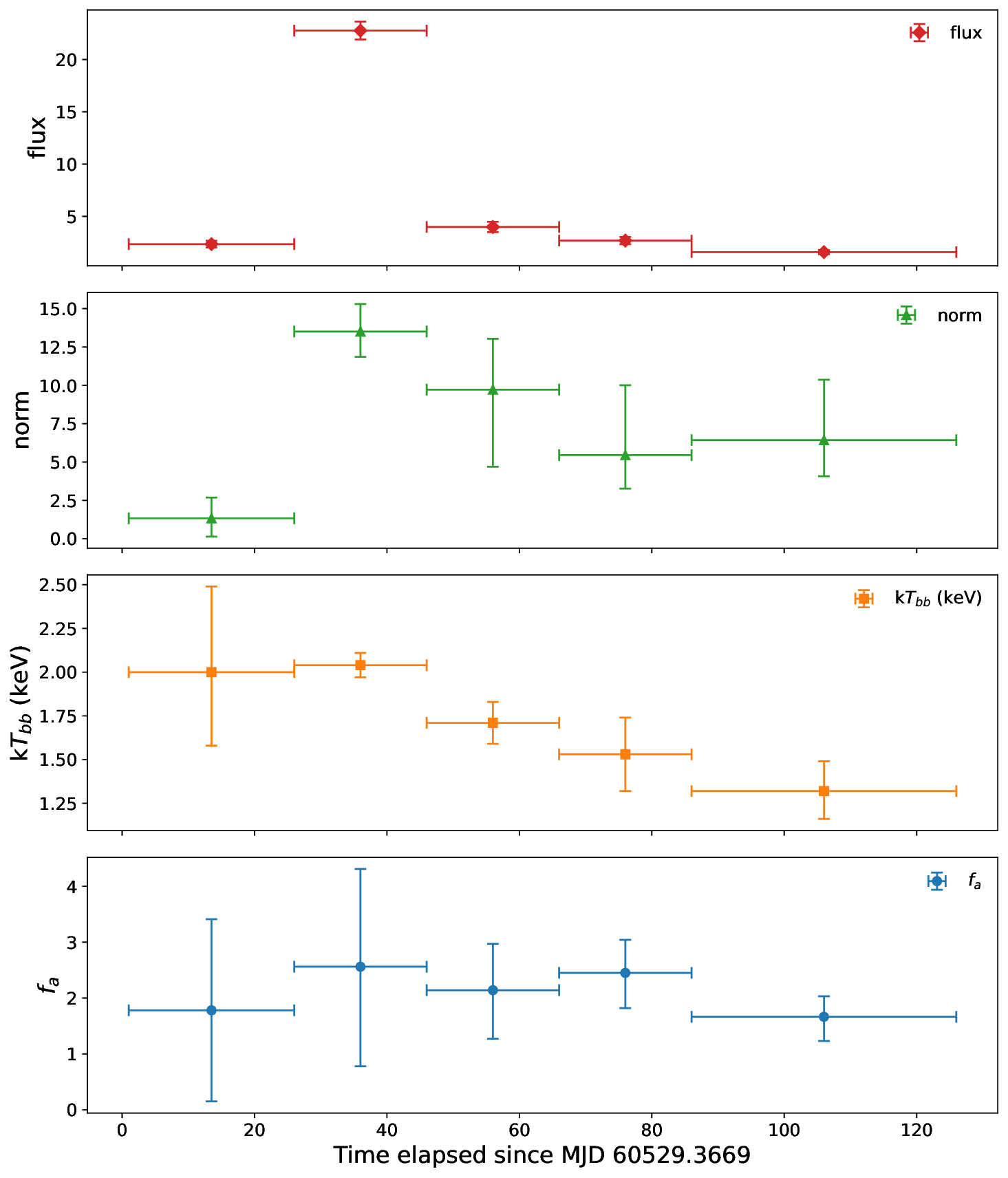}
		\includegraphics[width=0.3\columnwidth]{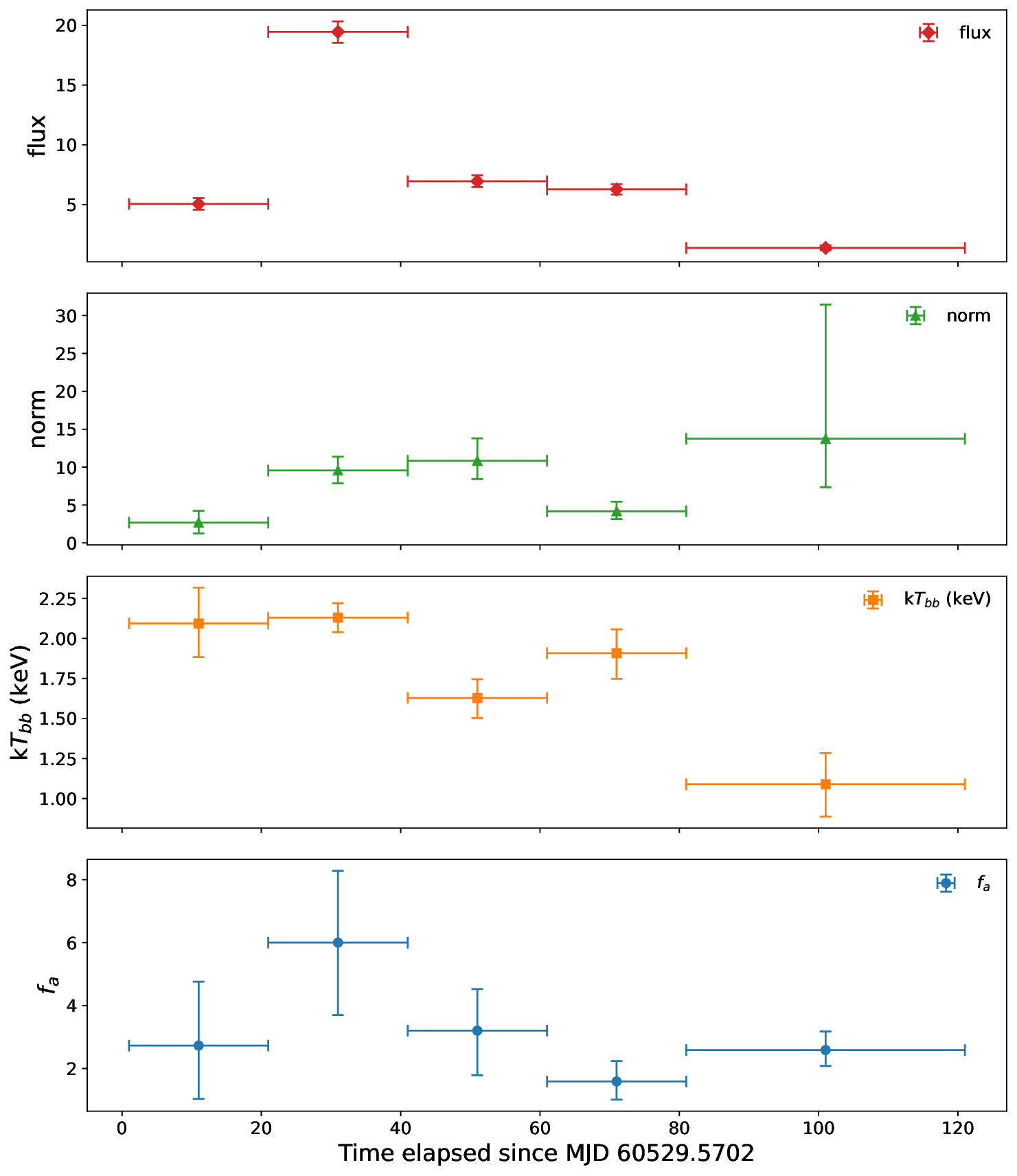}
		\includegraphics[width=0.3\columnwidth]{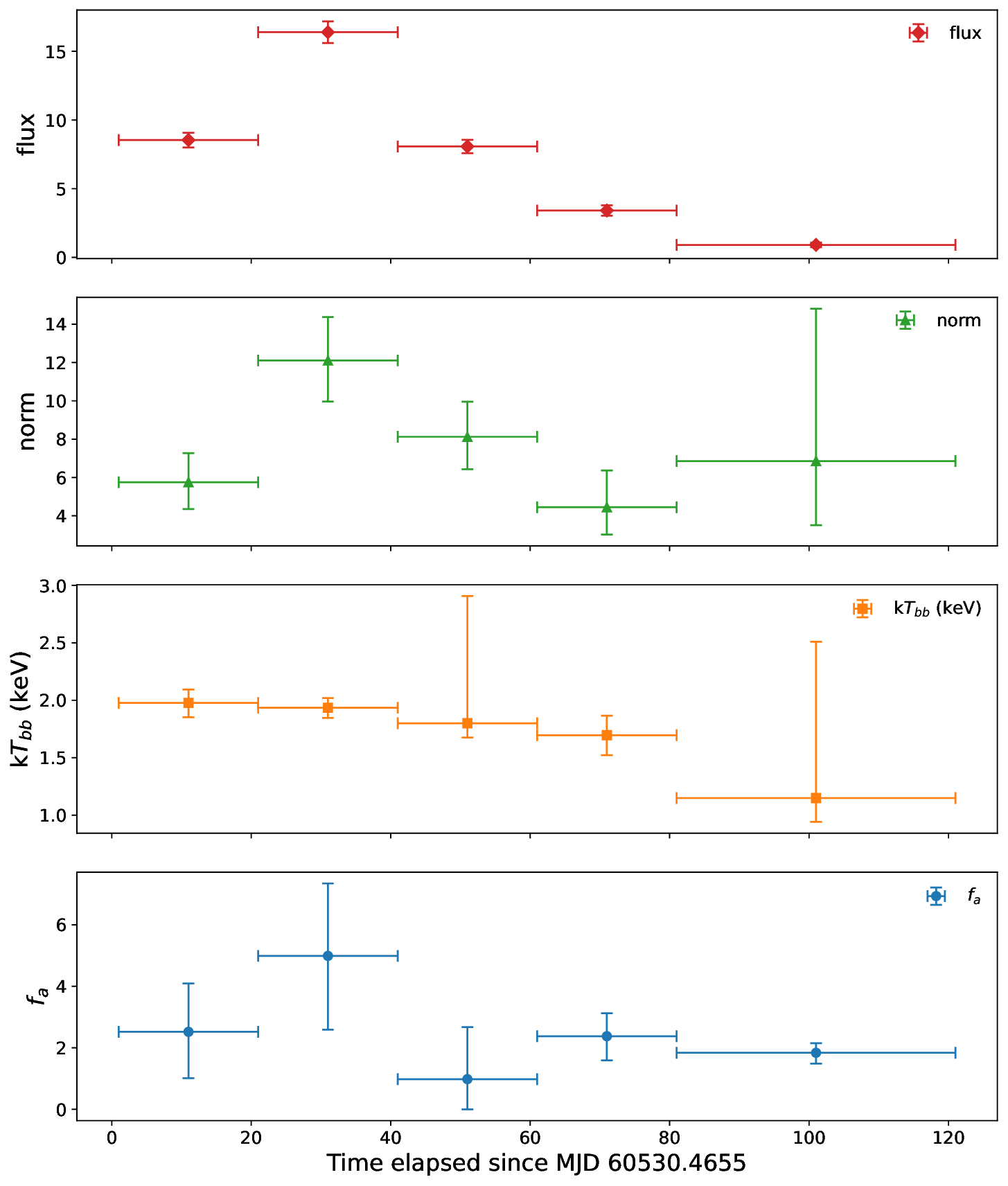}
		\caption{Evolution of the spectral parameters for blackbody emission during the bursts B1, B2 and B5 (ref. Table \ref{tab:Burst_times}) for the \textit{NuSTAR}/FPMA observation of the source 4U 1323-62. The best-fit parameters corresponding to the employed Model 7: \texttt{Tbabs*($f_{\rm a}$*edge*nthcomp+bbodyrad)} in the energy range 3-20 keV for - \textbf{left panel}: burst 1 (B1) corresponding to Table \ref{tab:fit_B1} for a duration of 125s; \textbf{middle panel}: burst 2 (B2) corresponding to Table \ref{tab:fit_B2} for a duration of 120s; and \textbf{right panel}: burst 5 (B5) corresponding to Table \ref{tab:fit_B5} for a duration of 120s. Within each panel, the plots from top shows- \textbf{\textit{first panel}}: plot of blackbody flux in units of $10^{-10}$ ergs cm $^{-2}$ s $^{-1}$; \textbf{\textit{second panel}}: plot of the blackbody norm; \textbf{\textit{third panel}}: plot of the temperature of the blackbody emission (k$T_{\rm bb}$); and \textbf{\textit{fourth panel}}: plot of the component for comparing the burst emission with the persistent emission before burst ($f_{\rm a}$), corresponding to the bursts B1, B2 and B5. Horizontal bars indicate the width of the time bins, and the vertical bars are $1\sigma$ uncertainties.}
		\label{fig:best_fit_par}
	\end{figure*}

	\quad Of the six bursts, we performed time-resolved spectral analysis for three- B1, B2, and B5-while the remaining bursts lacked sufficient data points or yielded poor fit values ($\chi^2$/dof). Each time-resolved spectrum was fitted with Model 7, which incorporates both the burst and enhanced persistent emissions. The evolution of blackbody temperature, flux, normalization, and the scaling factor ($f_{\rm a}$) for these bursts is shown in Fig. \ref{fig:best_fit_par}.
	
	In all three cases, the maximum blackbody flux occurs during segment S2, corresponding to the burst peak. The maximum peak fluxes (in units of $10^{-10}$ erg cm$^{-2}$ s$^{-1}$) are $22.76_{-0.86}^{+0.86}$ for B1, $19.45_{-0.91}^{+0.88}$ for B2 and $16.39_{-0.79}^{+0.79}$ for B5 indicating a gradual decline from B1 to B5. The normalization parameter (norm), which traces the apparent radius of the blackbody-emitting region, peaks at S2 for both B1 and B5, decreases to a minimum at S4, and increases slightly at S5 (Fig. \ref{fig:best_fit_par}). However, the \textit{norm} values at S5 for each burst carry large uncertainties—e.g., $13.74_{-6.40}^{+17.71}$ for B2 and $6.97_{-3.34}^{+7.96}$ for B5 (Table \ref{tab:fit_B2})—indicating limited reliability. The norm values at S2 ($9.55_{-1.70}^{+1.82}$) and S3 ($10.83_{-2.40}^{+2.97}$) in B2 correspond to blackbody radius of 3.09-3.29 km. The similarity of S2 and S3 values suggests that the norm for B2 shows a similar evolution pattern to B1 and B5. The maximum estimated blackbody radii are $3.67_{-0.23}^{+0.24}$ km for B1, $3.29_{-0.39}^{+0.42}$ km for B2, and $3.48_{-0.32}^{+0.31}$ km for B5.\\
	
	The evolution of the blackbody temperature (k$T_{\rm bb}$) for bursts B1, B2, and B5 (Fig. \ref{fig:best_fit_par}) shows a rise during the burst phase followed by a decline during decay, with the maximum temperature occurring in segment S2. Since the burst rise (S1) transitions into the peak at the S1-S2 boundary, the k$T_{\rm bb}$ values for these two segments are nearly identical. The peak temperatures are $2.04_{-0.07}^{+0.07}$ keV, $2.13_{-0.09}^{+0.09}$ keV, and $1.98_{-0.13}^{+0.12}$ keV for B1, B2, and B5, respectively (Tables \ref{fig:B1_b,f}, \ref{fig:B2_b,f}, and \ref{fig:B5_b,f}). The persistent flux ($F_{\rm f_{\rm a} * nthcomp}$) exceed the blackbody flux while approaching towards the final segment (S5) of the burst exposure as indicated by the best-fit spectral parameters for B1, B2 and B5 corresponding to Model 7 (see the lower panels of S5 in Fig. \ref{fig:B1_b,f}, \ref{fig:B2_b,f}, \ref{fig:B5_b,f}). This growing dominance of the persistent emission implies enhanced up-scattering of low-energy photons, consistent with the mechanism described by \citealt{1966ApJ...145..560W}.\\
	
	The scaling factor ($f_{\rm a}$) quantifies the enhancement of persistent emission relative to its pre-burst level. From the time-resolved spectral studies of the bursts with the $f_{\rm a}$ model, we found a significant enhancement of the persistent emission during the burst, quantified by $f_{\rm a}$. The strength of the persistent emission enhancement scales directly with the burst luminosity. For all three bursts, $f_{\rm a}$ reaches its maximum in segment S2, coinciding with the burst peak. The peak $f_{\rm a}$ values are $2.56^{+1.75}_{-1.78}$ for B1, $6.00_{-2.30}^{+2.28}$ for B2, and $4.99_{-2.39}^{+2.36}$ for B5. For B2, the most powerful burst amongst B1, B2, and B5 required a substantial enhancement ($f_{\rm a} \simeq 6$) of the persistent emission during the burst peak. This trend provides quantitative evidence for a physical scenario where the strength of the burst-disc interaction, likely caused by the Poynting-Robertson drag, is a direct function of the thermonuclear flash’s radiative power \citep{worpel2013evidence, Worpel_2015}.
	
	In addition, a 30-70 keV light curve has been plotted to provide another means of diagnosing the burst's influence on the accretion disk/corona. The light curve for all the bursts, aligned at their start times in both the 30-70 keV and 3-70 keV energy bands, are shown in the lower panel of Fig. \ref{fig:bursts_lc}, where a clear decrease in the count rate during burst interval is evident for all bursts. From the inset plot in the lower panel of Fig. \ref{fig:bursts_lc}, the	burst count rate in 30-70 keV hard energy band is observed to vary within 3-4 counts/s at the peak and decrease to less than 3 counts/s during the decay phase. In the 3-70 keV energy band, during the peak emission of the burst, the count rate exceeds 100 counts/s for all bursts except B1 and B4. During the decay phase of the bursts, the count rate decreases to 5-20 counts/s. The drop in the flux in the hard energy band is directly related to the hard X-ray deficits observed during the spectral analysis of type-I X-ray bursts and is considered a key signature of the burst-corona interaction. Thus, the hard X-ray deficit observed in the burst spectra can be explained by the burst-corona interaction (e.g., scattering of the burst photons by the corona). Moreover, during the bursts, the flux from both the thermal and Comptonized components increase, with the thermal photons serving as the seed photons for the Comptonization process. As demonstrated in earlier studies (e.g., \citealt{ji2014hard}), the substantial increase in soft-photon flux leads to enhanced Compton cooling of the corona, thereby reducing the coronal temperature. This observed hard X-ray deficit is a characteristic signature of Type-I thermonuclear bursts.\\

	\begin{table*}
		\centering
		\caption{Indicative parameters for the bursts calculated on the basis of the spectral parameters obtained corresponding to the best fit Model 7 (\texttt{Tbabs*($f_{\rm a}$*edge*nthcomp+bbodyrad)}) associated to the burst and pre-burst emission (spectra extracted from exposures of 600, 600, 1000, 300, 1000, and 1000 s, respectively, just before the onset of bursts B1, B2, B3, B4, B5, and B6) for the \textit{NuSTAR}/FPMA observation of 4U 1323-62 in the energy range 3-79 keV (if not stated in the respective column), where $\alpha$ is the ratio of the integrated persistent emission to the burst fluence, E$_{\rm burst}$ is the energy released during the burst emission, peak flux $F_{\rm peak}$, persistent flux $F_{\rm per}$, y$_{\rm ign}$ is the ignition depth and $Q_{\rm nuc}$ is the nuclear energy released for the material with solar abundances from spectral analysis.}
		\renewcommand{\arraystretch}{0.9}
		\resizebox{\textwidth}{!}{%
			\begin{tabular}{lccccccccc}
				\hline
				&&&&&&&&&\\
				\parbox{0.3cm}{Burst \\ Serial \\ no.} & \parbox{2.3cm}{\centering Bolometric flux$^\dagger$ \\ during the burst in \\ the energy range \\ $(0.1-100)$ keV} & \parbox{1.5cm}{\centering Burst fluence \\ $(f_{\rm b})^{\$}$} & \parbox{0.5cm}{$F_{\rm peak}^\dagger$} & \parbox{2.5cm}{\centering Decay time \\ of the burst \\ $\tau = f_{\rm b} /F_{\rm peak}$ \\ (s)} & \parbox{2.7cm}{\centering$E_{\rm burst} = 4 \pi d^2f_{\rm b}$ \\ $(10^{38}$ ergs $)$} & \parbox{0.4cm}{$F_{\rm per}^\dagger$} & \parbox{1.2cm}{\centering Ignition depth \\ ($y_{\rm ign}$)$^{\%}$} & \parbox{2.8cm}{$\alpha = \frac{c_{\rm bol} F_{\rm per} \triangle t_{\rm rec}} {{f_{\rm b}}}$}  & \parbox{0.5cm}{$Q_{\rm nuc}^{\#}$} \\
				
				\hline
				&&&&&&&&&\\
				B1 & 13.95 & 1.46 & 31.44 & 46.58 & 17.52 & 3.36 & 0.63 & 62.31 & 2.99 \\
				B2 & 13.61 & 1.46 & 38.89 & 37.43 & 17.42 & 3.03 & 0.56 & 56.57 & 3.30 \\
				B3 & 14.43 & 1.57 & 30.93 & 50.84 & 18.82 & 3.13 & 0.58 & 53.98 & 3.45 \\
				B4 & 10.70 & 0.39 & 13.05 & 29.52 & 4.61 & 3.09 & 0.58 & 217.52 & 0.86 \\
				B5 & 13.29 & 1.44 & 32.73 & 43.85 & 17.17 & 3.35 & 0.63 & 63.44 & 2.94 \\
				B6 & 12.91 & 1.18 & 30.35 & 38.72 & 14.06 & 3.22 & 0.60 & 74.46 & 2.50 \\
				
				\hline
				\textit{Note:} & \multicolumn{3}{l}{ $^{\dagger}$ denotes the unit of flux $10^{-10}$ erg cm $^{-2}$ s $^{-1}$;} 		 & \multicolumn{2}{l}{$^{\$}$ denotes the unit in $10^{-7}$  erg cm$^{-2}$;}
				& \multicolumn{3}{l}{$^{\%}$ denotes the unit of y$_{\rm ign}$ in $10^8$ g cm$^{-2}$;} & \\ & \multicolumn{9}{l}{$^{\#}$ denotes the unit of $Q_{\rm nuc}$ in $10^{18}$ erg g$^{-1}$.} \\
			\end{tabular}
		}
		\label{tab:spect_best_fit}
	\end{table*}
	
	Using the burst light curve data as listed in Table \ref{tab:Burst_times}, we have estimated the spectral parameters of the blackbody emission in Table \ref{tab:spect_best_fit}. The bolometric flux for each burst were obtained by applying the \texttt{flux} command to our spectral model (Model: 7 \texttt{Tbabs*($f_{\rm a}$*edge*nthcomp+bbodyrad)}) in \texttt{XSPEC} over the energy range 0.1 - 100 keV. Following \citealt{galloway2008thermonuclear}, a bolometric correction was applied since the persistent emission extends over a broader energy range ($>$ 80 keV) than the burst emission ($<$ 25 keV). As our observational data cover only 3–79 keV, we have used the \texttt{XSPEC} command \texttt{dummyrsp} that generates a response beyond the observed energy limit (\citealt{arnaud2003x}). The burst fluence is the measure of the total energy radiated during the burst, measured in g cm$^{-2}$(\citealt{GRBCAT}). The burst fluence ($f_{\rm b}$) obtained for the bolometric flux over the whole energy range ($0.1 - 100$ keV) during the burst integration times (Table \ref{tab:Burst_times}) are listed in Table \ref{tab:spect_best_fit}. Among the bursts, B1, B2 and B5 show fluences of 1.464, 1.456 and 1.435 (in units of 10$^{-7}$ erg cm$^{-2}$), respectively. The burst fluence does not follow a monotonic trend-it increases twice (in units of 10$^{-7}$ erg cm$^{-2}$): from 1.456 (B2) to 1.572 (B3), and again from 0.385 (B4) to 1.435 (B5). Typically, type-I thermonuclear bursts exhibit a gradual decrease in burst fluence as the neutron star surface depletes its nuclear fuel after each burst. \\
	
	The burst fuel composition can be inferred from Q$_{\rm nuc}$, which is the nuclear energy released per nucleon for material with solar abundances (\citealt{galloway2008thermonuclear}). For a neutron star with $M=1.4 M_{\rm \odot}$ and radius $R=10 {\rm km}$, Q$_{\rm nuc}$ (in units of 4.4 MeV nucleon$^{-1}$) is given by $Q_{\rm nuc} = \frac{44}{\alpha}$, where $\alpha$ is the ratio of the integrated persistent flux to the burst fluence. The value of $\alpha$ is given as $\alpha = \frac{c_{\rm bol} F_{\rm per} \triangle t_{\rm rec}} {f_{\rm b}}$, where c$_{\rm bol}$ is the bolometric correction parameter, $F_{\rm per}$ is the flux during persistent emission before the burst and $\triangle t_{\rm rec}$ is the recurrence time between consecutive bursts. While burst typically spans 2.5-25 keV, a broader energy range (upto $\sim$ 100 keV) must be considered for persistent emission due to its high-energy cutoff ($15-80$ keV). In \citealt{galloway2008thermonuclear}, spectra were modeled over 0.1-200 keV, and c$_{\rm bol}$ was calculated as a ratio between flux in the 0.1-200 keV and 2.5-25 keV energy bands. The derived bolometric correction factor ($c_{\rm bol}$) for the \textit{RXTE} observation of the source 4U 1323-62 was 1.67$\pm$0.05 (\citealt{galloway2008thermonuclear}). Next, the recurring times between bursts - B1-B2; B3-B4; and B5-B6 are estimated as 4.88 hr, 4.44 hr and 4.24 hr, respectively yielding an average of $\triangle t_{\rm rec}$ = $4.52\pm0.32$ hr. Earlier studies reported burst recurrence times of $\sim$5.30–5.43 hr with \textit{EXOSAT} (1985), $\sim$3.05 hr with \textit{ASCA} (1994), $\sim$2.40–2.57 hr with \textit{BeppoSAX} (1997), $\sim$2.45–2.59 hr with \textit{RXTE} (1997), $\sim$5.59-5.65 hr with \textit{Suzaku} (2007) and $\sim$2.65-2.70 hr with \textit{AstroSat} (2017) \citep{1989ApJ...338.1024P, balucinskachurch1999xraystudydippinglow, barnard2001, balucinska2009neutral, Bhulla_2020}. Persistent emission spectra were extracted in the 3–79 keV energy range from intervals preceding each burst, as described in Section \ref{subsec:pers_time}. The persistent fluxes ($F_{\rm per}$) were calculated using the \texttt{flux} command applied to the best-fit results of Model 6 (\texttt{Tbabs*(edge*nthcomp+bbodyrad)}). For the burst peak intervals (segment S2, corresponding to $>$90\% of the peak count rate), the peak fluxes ($F_{\rm peak}$) were similarly derived using the \texttt{flux} command on the best-fit results of Model 7 (\texttt{Tbabs($f_{\rm a}$*edge*nthcomp+bbodyrad)}). From Table \ref{tab:spect_best_fit}, we note $\alpha$ values for all bursts range between $\sim$ 54 and 74, except B4. According to \citealt{2021ASSL..461..209G}, such $\alpha$ values (40–100) are characteristic of bursts powered by mixed hydrogen/helium fuel.
	
	Except for B4, the burst decay times range from 37.4-50.8 s, classifying them as short bursts ($\leq$ few minutes). B4 shows distinct properties with $\alpha \sim 218$, the shortest decay time (29.5 s), and the lowest fluence (0.385 $\times10^{-7}$ erg cm$^{-2}$). Such a high $\alpha$ typically indicates either accretion-driven emission or incomplete fuel burning (\citealt{10.1093/mnras/staa2749}). In contrast, the subsequent bursts (B5 and B6) with $\alpha \sim$ 64–74 are consistent with mixed H/He-fueled events. 
	
	The large $\alpha$ of B4 suggests enhanced accretion energy and reduced nuclear burning efficiency, as following \citealt{galloway2008thermonuclear}, the parameter $\alpha$ can also be expressed as $\alpha = \frac{Q_{\rm grav}}{Q_{\rm nuc}}(1+z)$, where $Q_{\rm grav}$ is the gravitational potential energy released during accretion and z represents redshift which corresponds to $(1+z)=1.31$ for a neutron star with mass $M=1.4M_{\rm \odot}$ and radius R=10 km. Occurring at the bottom of an intensity dip, B4’s low flux likely results from coronal scattering (\citealt{1987A&A...178..137F}). Its peak count rate (94 counts s$^{-1}$) is 38\% below the mean (152 counts s$^{-1}$), and its peak flux is 55\% below the average, though its pre-burst persistent flux is not the lowest. The rise time (2 s) is also significantly shorter than in other bursts ($>5$ s). In this regard, \cite{jonker1998persistent} reported the detection of a secondary burst in two observations, as well as a burst occurring during one of the dips, in the 1997 \textit{RXTE} observation of the source. As noted by \cite{albayati2021discovery}, certain bursts can display multiple peaks, and such double-peaked profiles are sometimes interpreted as separate bursts (\citealt{bhattacharyya2007unusual}). Therefore, B4’s low peak and short rise time may indicate that it represents a double-peaked burst event. This interpretation is further supported by the fact that burst B3 shows the highest fluence ($\sim$1.572 $\times$ $10^{-10}$ erg cm$^{-2}$) and the largest total energy release ($\sim$1.882 $\times$ $10^{39}$ erg), suggesting significant fuel burning immediately before the weaker B4 event.\\
	
	The burst fluence represents the total radiated energy during a burst, expressed as (E$_{\rm burst}$) in units of $10^{38}$ erg, calculated using $E_{\rm burst} = 4 \pi d^2f_{\rm b}$, where the distance to the source ($d$) is assumed to be 10 kpc (\citealt{Bhulla_2020}). Except B4, all bursts yield E$_{\rm burst}$ values on the order of $10^{39}$ erg. The nuclear energy released for material with solar abundance is given by $Q_{\rm nuc}=1.6 + 4 \langle X \rangle$ (in units of 4.4 MeV nucleon$^{-1}$), where, $\langle X \rangle$ denotes the hydrogen mass fraction (\citealt{cumming2003models}). For a fuel composition containing 90\% hydrogen ($\langle X \rangle$ $\sim$ 0.9), the corresponding nuclear energy release is $Q_{\rm nuc} =$ 5.2 MeV/nucleon ($\simeq5.0\times10^{18}$ erg g$^{-1}$). As given in Table \ref{tab:spect_best_fit}, the value of $Q_{\rm nuc}$ (in units of $\times 10^{18}$ erg g$^{-1}$) attains a maximum value of 3.45 for B3 and remains above 2.5 for all bursts except B4. According to \cite{Galloway_2020}, such values indicate thermonuclear burning of mixed H/He fuel through rapid proton capture (rp) process, characteristic of short bursts. The comparatively low $Q_{\rm nuc}$ of B4 suggests predominant helium burning into heavier nuclei.\\
	
	\cite{1981ApJ...247..267F} examined hydrogen and helium burning on accreting neutron stars, developing models for pressure and temperature conditions necessary for unstable ignition at varying fuel-layer depths. Following \cite{galloway2008thermonuclear}, the ignition column depth ($y_{\rm ign}$) in units of 10$^8$g cm$^{-2}$ for a neutron star of radius $R_{\rm NS}$ is given as
	
	\begin{align}
		\nonumber
		y_{\rm ign} &= \frac{E_{\rm burst}(1+z)}{4 \pi R_{\rm NS}^2 Q_{\rm nuc}} \\
		\nonumber
		&= 3.0 \times 10^8 \left(\frac{f_{\rm b}}{10^{-6} \, ergs \, cm^{-2}}\right) \left(\frac{d}{10 \, kpc}\right)^2  \left(\frac{1+z}{1.31}\right) \left(\frac{R_{\rm NS}}{10 \, km}\right)^{-2} \\
		\nonumber
		&\times\left(\frac{Q_{\rm nuc}}{4.4 \, MeV \, nucleon^{-1}}\right)^{-1}
	\end{align}

	The computed ignition depths ($y_{\rm ign}$) are listed in Table \ref{tab:spect_best_fit} for bursts B1-B6 are approximately 0.63, 0.56, 0.58, 0.58, 0.63 and 0.60 $\times$$10^8$ g cm$^{-2}$, respectively. The identical ignition depths for B3 and B4 imply a comparable accreted mass column at those times. As the reported values are in the same order, we infer an average ignition depth $y_{\rm ign}^{\rm overall}\sim0.6\times10^8\,g\,cm^{-2}$. According to \cite{bildsten1997thermonuclearburningrapidlyaccreting} and \cite{cumming2000rotational}, such ignition depths are consistent with short Type-I thermonuclear bursts powered by mixed hydrogen/helium fuel.\\
	
	\section{Data Availability}
	We used the archival data from the \texttt{NASA}'s \texttt{HEASARC} database.
	
	\section{Acknowledgements}
	A note of thanks is extended to the esteemed reviewer for the valuable comments that have improved the quality of this manuscript. We thank \texttt{NuSTARDAS} for ensuring proper extraction of usable data. For the analysis of archival data we deeply acknowledge \texttt{HEASOFT} and \texttt{CALDB}. A token of thanks is extended towards the non-NET fellowship program of Visva-Bharati University for funding the research. We thank IUCAA for its vast repository of knowledge that enlightens curious minds.

	\def\apj{ApJ}
	\def\apjl{ApJl}
	\def\pasp{PASP} \def\mnras{MNRAS} \def\aap{A\&A} \def\physerp{PhR} \def\apjs{ApJS} \def\pasa{PASA}
	\def\pasj{PASJ} \def\nat{Nature} \def\memsai{MmSAI} \def\araa{ARAA} \def\iaucirc{IAUC} \def\aj{AJ} \def\aaps{A\&AS} \def\ssr{SSR}
	\def\iaucirc{iaucirc}
	\bibliographystyle{unsrtnat}
	\bibliography{4u1323m62_revised}
	
\end{document}